\documentclass{emulateapj}

\usepackage{color}
\usepackage{rotating}
\begin{document}
\title{Results from the Supernova Photometric Classification Challenge}

\email{kessler@kicp.uchicago.edu}
\submitted{accepted by PASP}

%

\newcommand{\NUMUCASTRO}{1}     
\newcommand{\NUMKICP}{2}        
\newcommand{\NUMCAPEMATH}{3}    
\newcommand{\NUMSAAO}{4}        
\newcommand{\NUMAIMS}{5}        
\newcommand{\NUMDESY}{6}        
\newcommand{\NUMDELHI}{7}       
\newcommand{\NUMPORT}{8}        
\newcommand{\NUMUCBoulder}{9}   
\newcommand{\NUMFNAL}{10}       
\newcommand{\NUMUTOR}{11}       
\newcommand{\NUMOXFORD}{12}     
\newcommand{\NUMRUTGERS}{13}    
\newcommand{\NUMANL}{14}        
\newcommand{\NUMGENEVE}{15}     
\newcommand{\NUMCALTECH}{16}    
\newcommand{\NUMSUSSEX}{17}     
\newcommand{\NUMKERALI}{18}     
\newcommand{\NUMCCC}{19}        
\newcommand{\NUMUCBASTRO}{20}   
\newcommand{\NUMUCBSTAT}{21}    
\newcommand{\NUMJHU}{22}        
\newcommand{\NUMUPENN}{23}      
\newcommand{\NUMPENNSTATE}{24}  
\newcommand{\NUMCAPEACGC}{25}   
\newcommand{\NUMCARNEGIE}{26}   
\newcommand{\NUMBOHR}{27}       
\newcommand{\NUMOSKAR}{28}      
\newcommand{\NUMCAPESTAT}{29}   


\author{
Richard~Kessler,\altaffilmark{\NUMUCASTRO,\NUMKICP}
Bruce~Bassett,\altaffilmark{\NUMCAPEMATH,\NUMSAAO,\NUMAIMS}
Pavel~Belov,\altaffilmark{\NUMDESY}
Vasudha~Bhatnagar,\altaffilmark{\NUMDELHI}
Heather~Campbell,\altaffilmark{\NUMPORT}
Alex~Conley,\altaffilmark{\NUMUCBoulder}
Joshua~A.~Frieman,\altaffilmark{\NUMUCASTRO,\NUMKICP,\NUMFNAL}
Alexandre~Glazov,\altaffilmark{\NUMDESY}
Santiago~{Gonz\'{a}lez-Gait\'{a}n},\altaffilmark{\NUMUTOR}
Ren\'{e}e~Hlozek,\altaffilmark{\NUMOXFORD}
Saurabh~Jha,\altaffilmark{\NUMRUTGERS}
Stephen~Kuhlmann,\altaffilmark{\NUMANL}
Martin~Kunz,\altaffilmark{\NUMGENEVE}
Hubert~Lampeitl,\altaffilmark{\NUMPORT}
Ashish~Mahabal,\altaffilmark{\NUMCALTECH}
James~Newling,\altaffilmark{\NUMCAPEMATH}
Robert~C.~Nichol,\altaffilmark{\NUMPORT}
David~Parkinson,\altaffilmark{\NUMSUSSEX}
Ninan~Sajeeth~Philip,\altaffilmark{\NUMKERALI}
Dovi~Poznanski,\altaffilmark{\NUMCCC,\NUMUCBASTRO}
Joseph~W.~Richards,\altaffilmark{\NUMUCBASTRO,\NUMUCBSTAT}
Steven~A.~Rodney,\altaffilmark{\NUMJHU}
Masao~Sako,\altaffilmark{\NUMUPENN}
Donald~P.~Schneider,\altaffilmark{\NUMPENNSTATE}
Mathew~Smith,\altaffilmark{\NUMCAPEACGC}
Maximilian~Stritzinger,\altaffilmark{\NUMCARNEGIE,\NUMBOHR,\NUMOSKAR}
and
Melvin~Varughese\altaffilmark{\NUMCAPESTAT}
} 


\altaffiltext{\NUMUCASTRO}{
Department of Astronomy and Astrophysics,
University of Chicago, 5640 South Ellis Avenue, Chicago, IL 60637
}

\altaffiltext{\NUMKICP}{
Kavli Institute for Cosmological Physics, 
University of Chicago, 5640 South Ellis Avenue Chicago, IL 60637
}

\altaffiltext{\NUMCAPEMATH}{
Department of Mathematics and Applied Mathematics, University
of Cape Town, Rondebosch 7701, South Africa.
}

\altaffiltext{\NUMSAAO}{
South African Astronomical Observatory, P.O. Box 9, Observatory
7935, South Africa.
}

\altaffiltext{\NUMAIMS}{
African Institute for Mathematical Sciences,
6-8 Melrose Road, Muizenberg 7945, South Africa
}

\altaffiltext{\NUMDESY}{
Deutsches Elektronensynchrotron (DESY) Notkestra{\ss}e 85, 
D--22607 Hamburg, Germany
}

\altaffiltext{\NUMDELHI}{
Department of Computer Science, University of Delhi, Delhi, India. 
}

\altaffiltext{\NUMPORT}{
Institute of Cosmology and Gravitation, 
Dennis Sciama Building, Burnaby Road,
University of Portsmouth, Portsmouth PO1 3FX, UK.
}

\altaffiltext{\NUMUCBoulder}{
Center for Astrophysics and Space Astronomy,
University of Colorado, Boulder, CO, 80309-0389.
}

\altaffiltext{\NUMFNAL}{
Center for Particle Astrophysics, 
Fermi National Accelerator Laboratory, P.O. Box 500, Batavia, IL 60510
}

\altaffiltext{\NUMUTOR}{
Department of Astronomy and Astrophysics, 
University of Toronto, 50 St. George Street, 
Toronto, ON, M5S 3H4, Canada
}

\altaffiltext{\NUMOXFORD}{
Department of Astrophysics, Oxford University, 
Oxford, OX1 3RH, UK.
}

\altaffiltext{\NUMRUTGERS}{
Department of Physics and Astronomy, 
Rutgers University, 136 Frelinghuysen Road, Piscataway, NJ 08854
}

\altaffiltext{\NUMANL}{
Argonne National Laboratory, 9700 South Cass Avenue, Lemont, IL 60437
}

\altaffiltext{\NUMGENEVE}{
D\'{e}partement de Physique Th\'{e}orique, Universit\'{e} de
Gen\`{e}ve, 24 quai Ernest Ansermet, CH1211 Gen\`{e}ve 4, Switzerland
}

\altaffiltext{\NUMCALTECH}{
California Institute of Technology, 
MC 249-17, 1200 East California Boulevard., Pasadena, CA 91125.
}

\altaffiltext{\NUMSUSSEX}{
Astronomy Centre, University of Sussex, Brighton BN1 9QH, UK.
}

\altaffiltext{\NUMKERALI}{
Department of Physics, St. Thomas College, 
Kozhencheri-689641 Kerala, India
}

\altaffiltext{\NUMCCC}{
Computational Cosmology Center, Computer Science Division, 
Lawrence Berkeley National Laboratory, 
1 Cyclotron Road, Berkeley, CA 94720.
}

\altaffiltext{\NUMUCBASTRO}{
Astronomy Department, 601 Campbell Hall, 
University of California Berkeley, Berkeley, CA 94720.
}

\altaffiltext{\NUMUCBSTAT}{
Statistics Department, 367 Evans Hall, 
University of California Berkeley, Berkeley, CA, 94720.
}

\altaffiltext{\NUMJHU}{
Department of Physics and Astronomy, 
Johns Hopkins University, Baltimore, MD 21218.
}

\altaffiltext{\NUMUPENN}{
Department of Physics and Astronomy,
University of Pennsylvania, 203 South 33rd Street, Philadelphia, PA  19104
}

\altaffiltext{\NUMPENNSTATE}{
Department of Astronomy and Astrophysics,
Pennsylvania State University,
525 Davey Laboratory, University Park, PA 16802.
}

\altaffiltext{\NUMCAPEACGC}{
Astrophysics, Cosmology and Gravity Centre (ACGC), Department
of Mathematics and Applied Mathematics, 
University of Cape Town, Rondebosch, 7701, South Africa.
}

\altaffiltext{\NUMCARNEGIE}{
Carnegie Observatories, Las Campanas Observatory,
Casilla 601, La Serena, Chile.
}

\altaffiltext{\NUMBOHR}{
Dark Cosmology Centre, Niels Bohr Institute, University
of Copenhagen, Juliane Maries Vej 30, 2100 Copenhagen \O, Denmark.
}

\altaffiltext{\NUMOSKAR}{
Department of Astronomy, Oskar Klein Centre,  
Stockholm University, 10691 Stockholm, Sweden.
}

\altaffiltext{\NUMCAPESTAT}{
Department of Statistical Sciences, University of Cape Town,
Rondebosch 7701, South Africa.
}



\newcommand{\DUEDATE}{May 1, 2010}
\newcommand{\SNPCC}{{\tt SNPhotCC}}
\newcommand{\SNPCCHOSTZ}{{\tt SNPhotCC/HOSTZ}}
\newcommand{\SNPCCnoHOSTZ}{{\tt SNPhotCC/noHOSTZ}}
\newcommand{\SNANA}{{\tt SNANA}}
\newcommand{\SDSS}{SDSS--II}

\newcommand{\unc}{uncertainty}
\newcommand{\uncs}{uncertainties}
\newcommand{\id}{identification}
\newcommand{\phot}{photometric}

\newcommand{\mlcs}{{\sc mlcs}}
\newcommand{\mlcsU}{{\sc mlcs-u2}}
\newcommand{\SALTII}{{\sc salt-ii}}
\newcommand{\SIFTO}{{\sc sifto}}

\newcommand{\Trestobs}{T_{\rm rest,obs}}
\newcommand{\Trest}{T_{\rm rest}}
\newcommand{\Tobs}{T_{\rm obs}}

\newcommand{\specy}{spectroscopically}
\newcommand{\spec}{spectroscopic}
\newcommand{\Specy}{Spectroscopically}
\newcommand{\Spec}{Spectroscopic}
\newcommand{\eff}{efficiency}
\newcommand{\ineff}{inefficiency}
\newcommand{\obss}{observations}
\newcommand{\obs}{observation}

\newcommand{\OM}{\Omega_{\rm M}}
\newcommand{\OL}{\Omega_{\Lambda}}

\newcommand{\wwwSDSS}{\tt http://www.sdss.org/}
\newcommand{\wwwPanSTARRS}{\tt http://pan-starrs.ifa.hawaii.edu/public}

\newcommand{\wwwNugent}{\tt http://supernova.lbl.gov/\~{}nugent/nugent\_templates.html}
\newcommand{\wwwSNANA}{\tt http://www.sdss.org/supernova/SNANA.html}
\newcommand{\wwwJEDI}{\tt http://jedi.saao.ac.za}
\newcommand{\wwwINCA}{\tt http://www.incagroup.org}
\newcommand{\wwwSIMGEN}{\tt http://sdssdp62.fnal.gov/sdsssn/SIMGEN\_PUBLIC}

\newcommand{\rspeclimit}{21.5}
\newcommand{\ispeclimit}{23.5}
\newcommand{\wwwdownload}{\tt www.hep.anl.gov/SNchallenge}
\newcommand{\photoz}{photo-$z$}

\newcommand{\NIaTOT}{{\cal N}_{\rm Ia}^{\rm TOT}}
\newcommand{\NIatrue}{N_{\rm Ia}^{\rm true}}
\newcommand{\NIafalse}{N_{\rm Ia}^{\rm false}}
\newcommand{\wIafalse}{W_{\rm Ia}^{\rm false}}
\newcommand{\wIafalseVALUE}{3}
\newcommand{\effIa}{\epsilon_{\rm Ia}}
\newcommand{\PurityIa}{PP_{\rm Ia}}
\newcommand{\effspec}{\epsilon_{\rm spec}}
\newcommand{\FoMIa}{{\cal C}_{\rm FoM-Ia}}
\newcommand{\FoMII}{{\cal C}_{\rm FoM-II}}

\newcommand{\curvedefSNIa}{
The first panel labeled ``All Ia tag''
is an arbitrary reference in which every SN has been
tagged as an SN~Ia, thereby ensuring 100\% efficiency.
The solid curves show $\pm 1\sigma_{\rm stat}$ 
values for the \specy\ confirmed subset,
and the dashed curves are for the unconfirmed SNe.
Entries are arranged by method categories 1--4
(\S\ref{sec:take_challenge}) 
as indicated in parentheses under the participant names
in the first panel.
}

\newcommand{\curvedefSNII}{
The first panel labeled ``All II tag''
is an arbitrary reference in which every SN has been
tagged as a type II, thereby ensuring 100\% efficiency.
The solid curves show $\pm 1\sigma_{\rm stat}$ 
values for the \specy\ confirmed subset,
and the dashed curves are for the unconfirmed SNe.
}


\newcommand{\NGROUP}{10}
\newcommand{\NSUBMIT}{22}
\newcommand{\NSUBMIThostz}{13}
\newcommand{\NSUBMITnohostz}{9}

\newcommand{\Trise}{T_{\rm rise}}
\newcommand{\Tfall}{T_{\rm fall}}
\newcommand{\nonIa}{non--Ia}
\newcommand{\zobs}{z_{\rm obs}}

\newcommand{\Ibc}{Ibc}
\newcommand{\IIn}{IIn}
\newcommand{\IIP}{II-P}
\newcommand{\IIL}{II-L}

\newcommand{\NumIbc}{16}
\newcommand{\NumIIn}{2}
\newcommand{\NumIIP}{23}
\newcommand{\NumIIL}{0}
\newcommand{\NumNONIa}{41}

\newcommand{\NSPEC}{1256}

\newcommand{\ProbIbc}{0.29}  
\newcommand{\ProbIIn}{0.04}
\newcommand{\ProbIIP}{0.59}
\newcommand{\ProbIIL}{0.08}

\newcommand{\NGENIa}{8000}  
\newcommand{\NCUTIa}{5300}  

\newcommand{\NGENnonIa}{9.3\times 10^{4}}
\newcommand{\NCUTnonIa}{1.3\times 10^{4}}

\newcommand{\magmin}{M_{\rm min}}
\newcommand{\maglim}{m_{\rm lim}}
\newcommand{\magpeak}{m_{\rm peak}}

\newcommand{\DZPHOT}{ (z_{\rm phot}-z_{\rm gen})/(1+z_{\rm gen}) }

\newcommand{\BESTFOM}{0.54}     
\newcommand{\BESTEFF}{0.96}     
\newcommand{\BESTPURITY}{0.79}  
\newcommand{\BESTPP}{0.56}      

\begin{abstract}
We report results from the 
Supernova Photometric Classification Challenge ({\SNPCC}),
a publicly released mix of simulated supernovae (SNe),
with types (Ia, \Ibc, and II) selected in proportion to 
their expected rate. The simulation was realized in the 
$griz$ filters of the Dark Energy Survey (DES)
with realistic observing conditions 
(sky noise, point-spread function and atmospheric transparency) 
based on years of recorded conditions at the DES site.
Simulations of \nonIa\ type SNe are based on 
\specy\ confirmed light curves that include 
{\it unpublished} \nonIa\ samples donated from the 
Carnegie Supernova Project (CSP), the
Supernova Legacy Survey (SNLS), and the
Sloan Digital Sky Survey-II (SDSS--II).
A \specy\ confirmed subset was provided for training.
We challenged scientists to run their classification
algorithms and report a type and \photoz\ for each SN.
Participants from {\NGROUP} groups contributed 
\NSUBMIThostz\ entries for the sample that included
a host-galaxy \photoz\ for each SN
and \NSUBMITnohostz\ entries for the sample that
had no redshift information.
Several different classification strategies resulted 
in similar performance, and for all entries the 
performance was significantly better for the training 
subset than for the unconfirmed sample.
For the \specy\ unconfirmed subset, 
the entry with the highest average
figure of merit for classifying SNe~Ia has an 
efficiency of \BESTEFF\ and an SN~Ia purity of \BESTPURITY.
As a public resource for the future development of photometric 
SN classification and \photoz\ estimators,
we have released updated simulations with improvements
based on our experience from the \SNPCC,
added samples corresponding to the 
Large Synoptic Survey Telescope (LSST) and the \SDSS,
and provided the answer keys so that developers can 
evaluate their own analysis.
\vspace{.5cm}
\end{abstract}

 \section{Motivation}
 \label{sec:intro}

To explore the expansion history of the universe, 
increasingly large samples of high-quality SN~Ia 
light curves are being used to measure luminosity 
distances as a function of redshift.  
With rapidly increasing sample sizes, there are 
not nearly enough resources to \specy\ confirm each SN.
Currently, the largest samples are from the 
Supernova Legacy Survey (SNLS: \cite{Astier06}) and the
Sloan Digital Sky Survey-II (SDSS-II: \citet{York00,Frieman07}),
each with more than 1000 SNe~Ia, 
yet less than half of their SNe are \specy\ confirmed.
The numbers of SNe are expected to increase dramatically
in the coming decade:
thousands for the Dark Energy Survey (DES: \citet{DES-moriond2009}) 
and a few hundred thousand for the 
Panoramic Survey Telescope and Rapid Response System
(Pan-STARRS)\footnote{\wwwPanSTARRS}
and the Large Synoptic Survey Telescope
(LSST: \citet{Ivezic_08,LSSTSB09}).
Since only a small fraction of these SNe will be \specy\ confirmed,
\phot\ \id\ is crucial to fully exploit these large samples.

In the discovery phase of accelerated cosmological expansion,
results were based on tens of high-redshift SNe~Ia,
and some samples included a significant fraction of events 
that were not classified from a spectrum
\citep{Riess_1998,Riess_2004,Perl_1997,Tonry_2003}.
While human judgment played a significant role in classifying
these ``photometric'' SNe,
more formal methods of \phot\ classification have been
developed over the past decade:
\citet{Poz2002,Poz2007a_classify,Dahlen2002,GalYam2004,Sulli2006,Johnson2006,Kuz2007,BEAMS2007,Rodney2009}.
Some of these techniques have been used to select candidates
for \spec\ observations and rate measurements 
\citep{Barris2006,Neill2006,Poz2007b_rate,Kuz2008,Dilday2008},
but these methods have not been used to select a significant 
\phot\ SN~Ia sample for a Hubble-diagram analysis.
In short, cosmological parameter estimates
from the much larger recent surveys are based
solely on \specy\ confirmed SNe~Ia
(SNLS: \citet{Astier06}, ESSENCE: \citet{WV07},
CSP: \citet{Freedman_2009}, SDSS-II: \citet{K09}).

The main reason for the current reliance on \spec\ \id\
is that vastly increased \spec\ resources have been used in 
these more recent surveys. 
In spite of these increased resources, however,
more than half of the discovered SNe 
lack \spec\ \obss, and therefore \phot\ methods must be used
to classify the majority of the SNe.
There are two difficulties limiting the application 
of \phot\ classification.
First is the lack of adequate \nonIa\ data for training algorithms.
Many classification algorithms were developed using
publicly available Nugent templates,\footnote{\wwwNugent}
consisting of a single spectral energy distribution
(SED) template for each \nonIa\ type.
The Nugent templates were constructed from averaging and 
interpolating a limited amount of \specy\ confirmed \nonIa\ data
\citep{Levan2005,Hamuy2002,Gill99,Baron2004,Cap1997},
and therefore the impact of the \nonIa\ diversity has 
not been well studied.
The second difficulty is that there is no standard 
testing procedure, and therefore it is not
clear which classification methods work best.

To aid in the transition to using \phot\ SN classification,
we have released a public 
``SN  Photometric Classification Challenge,'' 
hereafter called {\SNPCC}.
The announcement of the challenge and instructions to participants 
were given in a challenge release note \citep{SNchallenge2010},
and an electronic mail message alert was sent to several dozen
SN experts.
The \SNPCC\ consisted of a blinded mix of simulated SNe,
with types (Ia, Ib, Ic, II) selected in proportion to 
their expected rate.
From 2010 January 29 through June 1, the public challenge 
was open for scientists to run their classification
algorithms and report a type for each SN.
A \specy\ confirmed subset was provided so that algorithms
could be tuned with a realistic training set.
The goals of this challenge were to 
(1) learn the relative strengths and weaknesses of the different 
classification algorithms, 
(2) improve the algorithms, 
(3) understand what \specy\ confirmed subsets are needed to
properly train these algorithms, 
and (4) improve the simulations.

To address the paucity of \nonIa\ data, the CSP, SNLS and SDSS-II
generously contributed {\it unpublished} \specy\ 
confirmed \nonIa\ light curves.
These data are high-quality multiband light curves,
and we are grateful to the donating collaborations.
Since these \nonIa\ SNe are from surveys focused mainly 
on collecting type Ia SNe, this sample is brighter than 
the true \nonIa\ population.
In spite of this bias toward brighter \nonIa,
we anticipated that this challenge would
be a useful step away from the overly simplistic
studies that have relied on a handful of \nonIa\ templates.

The outline of this article is as follows.
In \S\ref{sec:sim} we present full details of the simulation,
including strengths, weaknesses and bugs found during the
\SNPCC.
In \S\ref{sec:take_challenge} we describe the classification
methods used by the \NGROUP\ participating groups.
The figure of merit used for evaluation is defined in 
\S\ref{sec:eval}, and the results for all of the \SNPCC\ 
participants are presented in \S\ref{sec:results}.
Updated simulations are described in \S\ref{sec:sim_update},
and we conclude in \S\ref{sec:end}.

 \section{The Simulation}
 \label{sec:sim}

Here we present full details of how the simulated
samples were generated using the
\SNANA\ software package\footnote{\wwwSNANA} \citep{SNANA09}.
Both the strengths and weaknesses are discussed
to motivate improvements in future simulations.
The limited information available to participants 
during the challenge is given in \S~2 of 
the challenge release note \citep{SNchallenge2010}.

 \subsection{Simulation Overview}
 \label{subsec:sim_overview}

The simulation was realized in the $griz$ filters of the 
Dark Energy Survey (DES), and distances were calculated
assuming a standard $\Lambda$CDM cosmology with
$\OM=0.3$, $\OL=0.7$ and $w=-1$.
The sky-noise, point-spread function and atmospheric transparency 
were evaluated in each filter
and each epoch using a year long history of actual conditions
from the ESSENCE project at the 
Cerro Tololo Inter-American Observatory 
(CTIO).\footnote{
The CTIO history of observing conditions is available in
the public \SNANA\ package (previous footnote).
} 
For the five SN fields selected for the DES (3 deg$^2$ per field), 
the cadence was based on allocating 10\% of the DES 
photometric observing time
and most of the nonphotometric time.
The cadence used in this publicly available simulation was 
generated by the Supernova Working Group within the 
DES collaboration.\footnote{Although two of us (RK \& SK) 
are members of the DES,
we did not include other DES colleagues in any discussions
about preparing the challenge, and we made our best efforts to 
prevent our DES collaborators from obtaining additional 
information beyond that contained in the release note.
} 
Since the DES plans to collect data during five months of the year,
incomplete light curves from temporal edge effects are included;
i.e., the simulated explosion times extend well before the start 
of each survey season, and extend well beyond the end of the season.

The \SNPCC\ included a sample with a host-galaxy 
photometric redshift ({\SNPCCHOSTZ})
and another sample with no redshift information ({\SNPCCnoHOSTZ}).
For the former, the \photoz\ estimates were based 
on simulated galaxies (for DES) analyzed with the
methods in \citet{Oy08a,Oy08b}.
The average host-galaxy \photoz\ resolution is 0.03,
and the \photoz\ distribution includes non-Gaussian outliers.
A challenge with precise \spec\ redshifts was not given
because using accurate redshifts makes little difference 
on the classifications compared with using a host-galaxy \photoz.

Two simple selection criteria were applied.
First, each object must have an observation in two or
more passbands with a signal-to-noise ratio (S/N) above 5.
Second, there must be at least five observations after explosion,
and there is no S/N requirement on these observations.
These requirements are relatively loose because part of 
the challenge was to determine the optimal selection criteria.
For the five seasons planned for the DES, the total number of 
generated SNe for all types  was $1.01\times 10^5$. 
The number satisfying the loose selection requirements 
and included in the \SNPCC\  was $1.8\times 10^4$.

 \subsection{Type Ia Model}
 \label{subsec:simIa}

Simulated SNe~Ia were  based on an equal mix of the
\SALTII\ \citep{Guy07} and \mlcs\ models \citep{JRK07,K09}. 
Since these two models do not agree in the ultraviolet region,
we used a special {\mlcsU} version in which the
ultraviolet region was adjusted to match that of the
\SALTII\ model. 
The treatment of color variations corresponding to 
each model was used.
For \mlcsU, extinction by dust resulted in reddened SNe~Ia.
The dust parameter $R_V$ was drawn from an asymmetric Gaussian
distribution peaked at $R_V=2.0$ with sigmas of 0.2 and 0.5 for the
low and high side, respectively, and the $R_V$ values were
constrained to lie between 1.5 and 4.1; this $R_V$ distribution
has a mean value of 2.2.
For \SALTII, the color-magnitude adjustment was given by $\beta c$
where $\beta = 2.7$ and $c$ is the color excess, $E(B-V)$.
The $c$ parameter was drawn from a Gaussian distribution
with $\sigma_c = 0.1$ and the constraint
$\vert c \vert < 0.4$.

In addition to the model parameters, we have simulated the 
anomalous Hubble scatter with random color variations.
For each passband $f$, a random shift was drawn from a
Gaussian distribution with $\sigma_m = 0.09$~mag,
and this magnitude shift was applied coherently to
all epochs within the passband. The scatter in each
color was therefore $0.09\cdot \sqrt{2}$~mag.

For the \SNPCCnoHOSTZ\ it is important to include
photometric passbands that correspond to rest-frame 
wavelengths outside the nominally defined ranges of the 
SN~Ia models: specifically, the $g$ and $r$ bands at 
higher redshifts that probe the far ultraviolet region.
Without an estimate of the redshift, analysis programs 
cannot initially select observations that correspond to a 
particular rest-frame wavelength range.
Since the spectral surfaces of the SN~Ia models are defined 
over a much larger range than  that where the models 
are defined, it is straightforward to extend the wavelength 
range in the simulation.
For both models, the lower wavelength 
range\footnote{The default rest-frame wavelength ranges 
for \mlcs2k2\ and \SALTII\ 
are 3200-9500~{\AA} and 2900-7000~{\AA}, respectively.
} 
was reduced to 2500~\AA.
To simulate redder passbands for \SALTII, the upper range 
was extended from 7000~{\AA} to 8700~{\AA}.

 \subsection{Non-Ia SN Model}
 \label{subsec:sim_nonIa}

\begin{figure*}
\centering
\epsscale{0.74}  
\plottwo{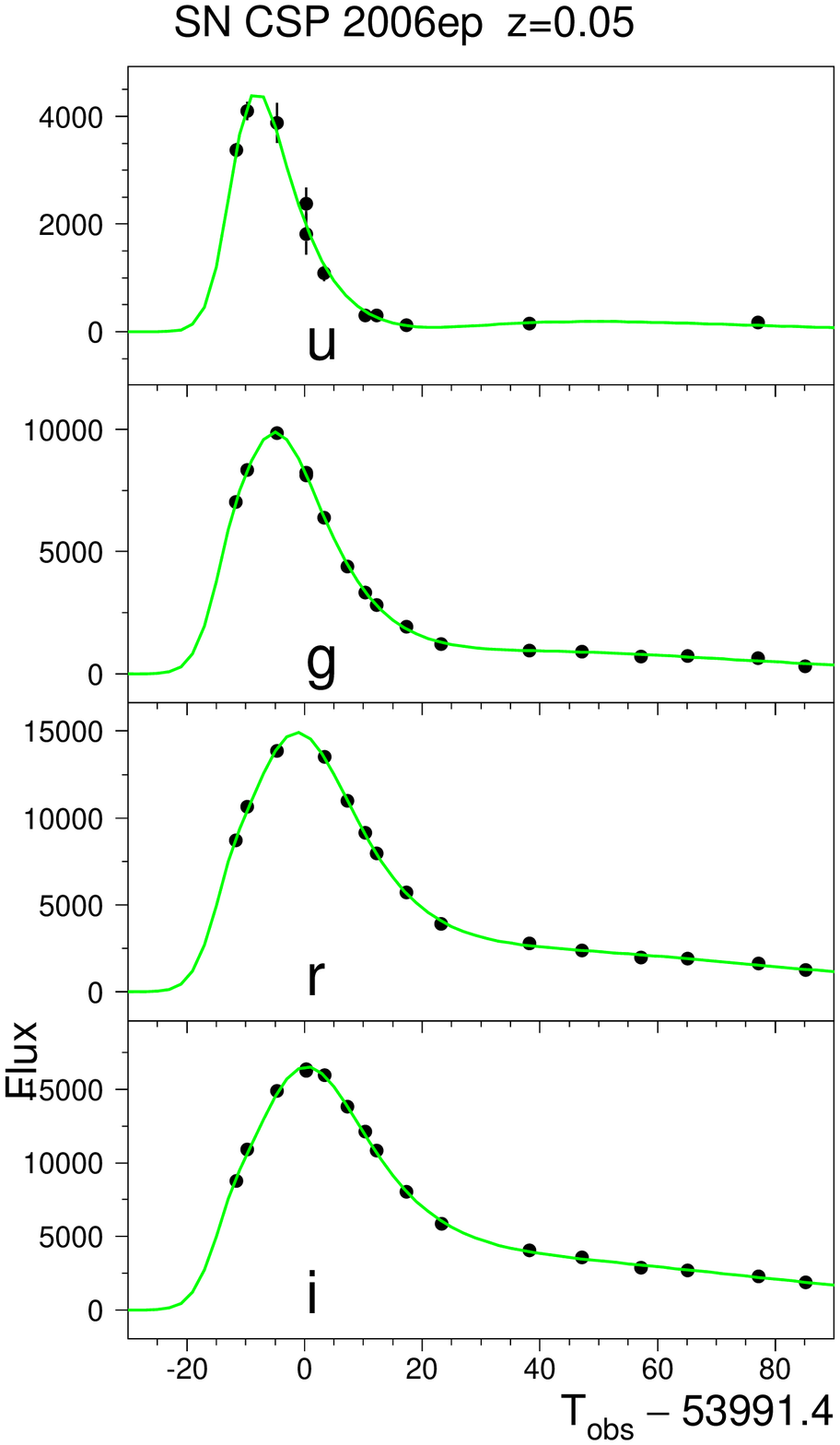}{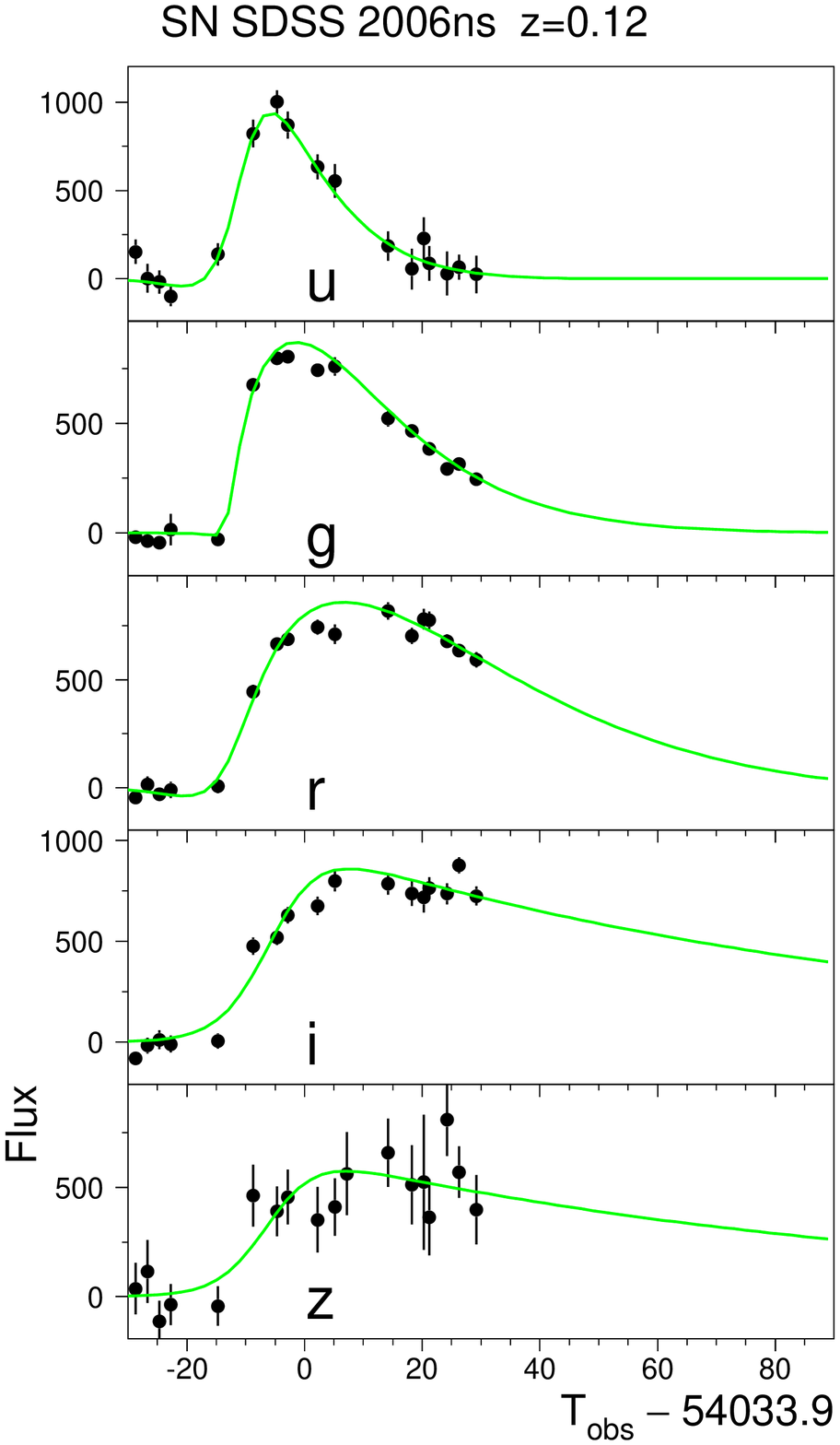}
\plottwo{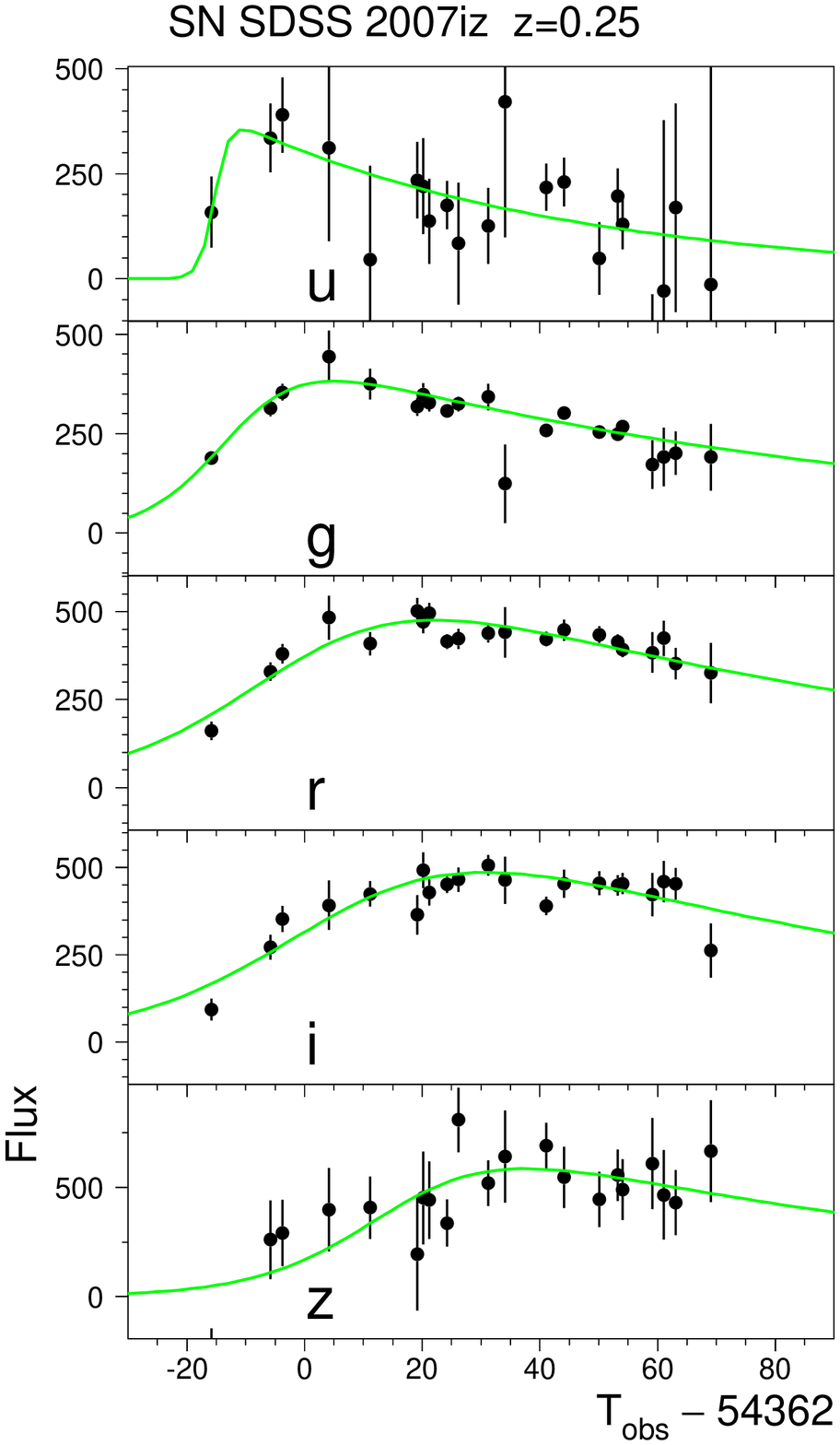}{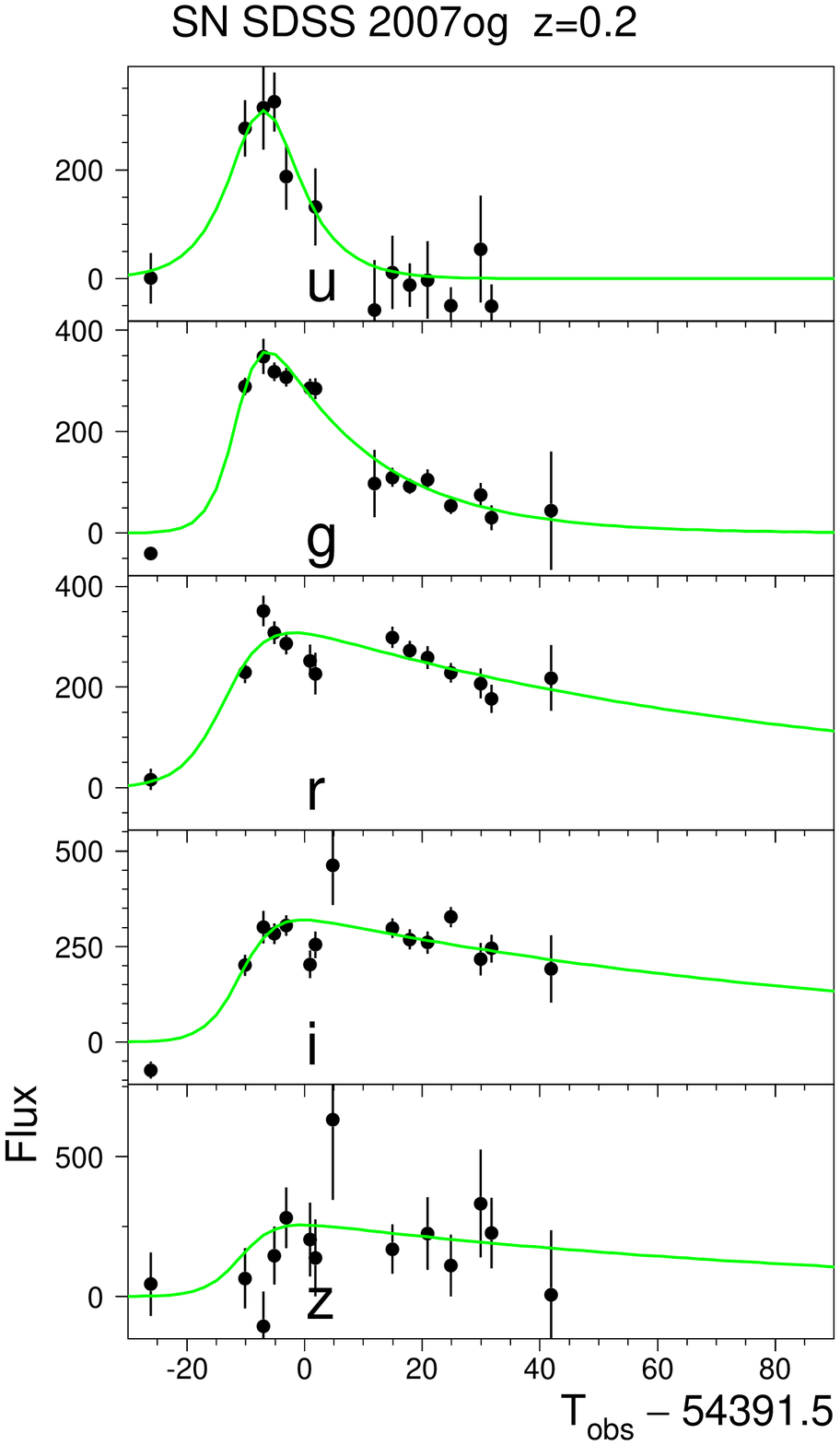}
\plottwo{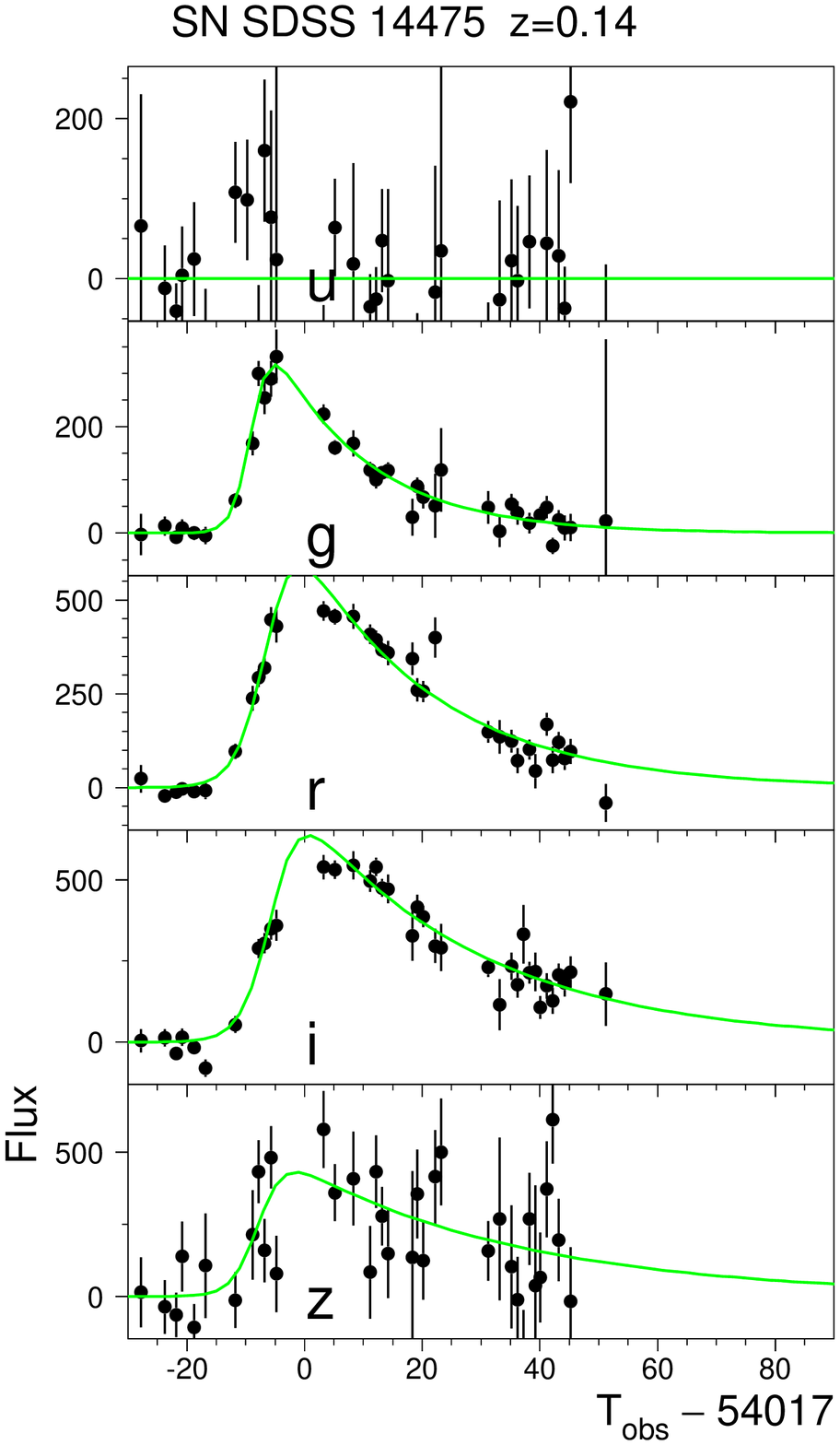}{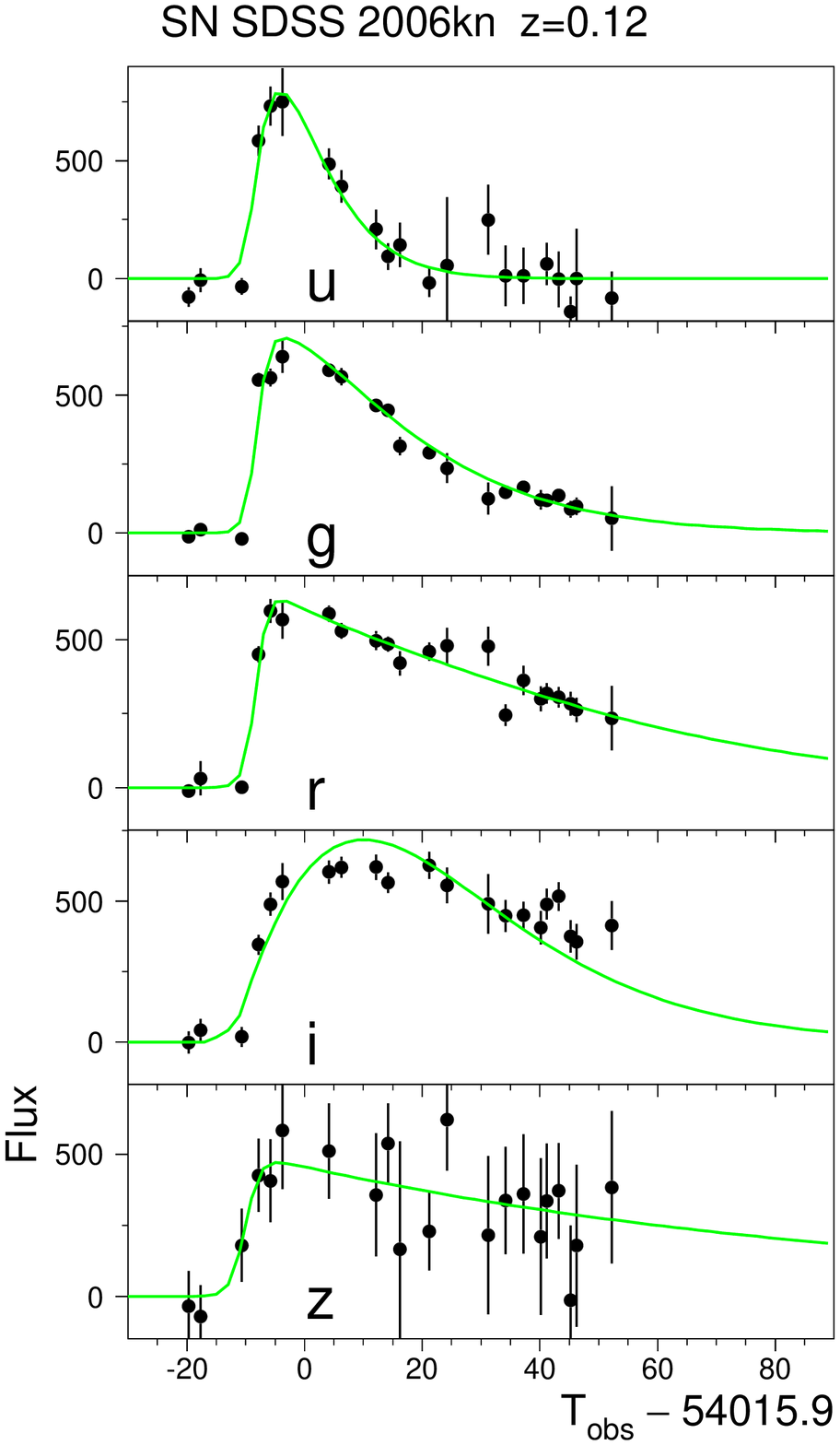} 
  \caption{
	\Specy\ confirmed \nonIa\ SNe data (black dots)
	resulting in the most misidentified \nonIa\ in the
	\SNPCC. The smoothing function in Eq.~\ref{eq:smooth}
	is shown by the green curve.
	The SN name and redshift are listed above each
	set of light curves.
	The filter is labeled in each panel.
      }
  \label{fig:smooth_non1a}
\end{figure*}

Simulated photometry of
\nonIa\ SNe was based on \specy\ confirmed \nonIa\
type light curves from the CSP, SNLS, and \SDSS\ SN surveys.
The basic strategy is to smoothly warp a standard SED to 
match the observed photometry and then use the warped SEDs
to simulate SNe at all redshifts.
After correcting the light curves for Galactic extinction,
the light curve for each passband was smoothed
using a general function based on that used in the \nonIa\ rate
analysis in \citet{Bazin2009},
\begin{equation}
  f(t) = A_0 [ 1 + a_1(t-t_0) +  a_2(t-t_0) ]
            \frac{    e^{-(t-t_0)/\Tfall} } 
                 {1 + e^{-(t-t_0)/\Trise} }~.
  \label{eq:smooth}
\end{equation}
The parameters $A_0$, $t_0$, $\Trise$, $\Tfall$ and $a_{1,2}$
are fit separately for each passband. The polynomial parameters
$a_{1,2}$ were initially fixed to zero; 
in cases where the fit was inadequate as determined by
visual inspection, the fit was redone with the
additional $a_{1,2}$ parameters.
Examples of smoothed light curves, also called \nonIa\ templates, 
are shown in Fig.~\ref{fig:smooth_non1a}
for the \nonIa\ SNe that were most commonly misidentified
as an SN~Ia during the \SNPCC\ (\S\ref{sec:results}).
To use a \nonIa\ template in the \SNPCC, 
the corresponding light curve was required to have
good sampling in all passbands, and this requirement 
was based on visual examination rather than rigorous cuts.
Among the 86 \specy\ confirmed \nonIa\ from the \SDSS,
34 were selected for the \SNPCC;
for the CSP, 5 of 6 were selected,
and for the SNLS, 2 of 9 were selected.
A list of the \NumNONIa\ \nonIa\ SNe used in the \SNPCC\ 
is shown in Table~\ref{tb:nonIa_list};
combining the surveys, the numbers of types \Ibc, \IIP\ and \IIn\  
are \NumIbc, \NumIIP, and \NumIIn, respectively
(also see Table~\ref{tb:nonIa_subtypes}).

While the general fitting function (Eq.~\ref{eq:smooth}) 
appears adequate upon visual inspection,
we note that the rise-time parametrization is not always accurate.
For SN 14475 in Fig.~\ref{fig:smooth_non1a}, the rise time
is well sampled and hence the smoothed template is reliable
in this region of the light curve.
For CSP-2006ep, however, the $u$-band rise time is not well sampled
and therefore the smoothed rise time is dependent on the
particular parametrization.  Ideally the rise-time shape
from well-measured \nonIa\ light curves would be used as
an additional constraint in the smoothing function,
but such constraints were not used in this \SNPCC.

The next step is to create a rest-frame time series of
SEDs such that the redshifted synthetic magnitudes match 
those of the smoothed light-curve template at each epoch.
These spectral time sequences are called
``\nonIa\ template SEDs.''
The starting SED for each \nonIa\ subtype is taken from 
the Nugent template,
and it is then warped at each epoch to match the
observer-frame photometry. 
For a simulated \nonIa\ type and redshift,
the corresponding \nonIa\ template SED is used to 
compute observer-frame $griz$ magnitudes.

In addition to the \NumNONIa\ \nonIa\ template SEDs
we have also included four Nugent SED templates,
each representing a composite average over one of the subtypes 
shown in Table~\ref{tb:nonIa_subtypes}.
The magnitudes were drawn from Gaussian distributions as 
described in \citet{Richardson2002}.

The final step is to apply random color variations
in the same manor as for the type Ia SNe.
While the anomalous scatter in the SN~Ia Hubble diagram 
motivates this step in the SN~Ia simulation, 
the motivation for the \nonIa\ simulation is to describe a 
potentially broader class of objects.
In the limit of a large and complete set of \nonIa\ 
templates there would be no need to simulate 
additional sources of magnitude variation.
We have made the assumption, however, that our set of \NumNONIa\ 
templates is not large enough to describe the 
{\nonIa} population.

\begin{table}
\caption{
	\Specy\ Confirmed Non-Ia SNe Used for Templates
  } 
\begin{center}
\begin{tabular}{llccc}
\tableline\tableline
        &    Spec  &  Observed  & \SNPCC\       & in            \\ 
 SN id  &    type  &  redshift  &  index\footnotemark[1]  & D10\footnotemark[2]   \\ 
\tableline 
 CSP 2004fe   &  Ic  & 0.0179 & 05   &  \\  
 CSP 2004gq   &  Ic  & 0.0055 & 06   &  \\  
 CSP 2004gv   &  Ic  & 0.0199 & 07   &  \\  
 CSP 2006ep   &  Ic  & 0.0495 & 08   &  \\  
 CSP 2007Y   &  Ic  & 0.0046 & 09   &  \\  
  \hline 
 SNLS 04D1la   &  \Ibc\  & 0.3190 & 10   &  \\  
 SNLS 04D4jv   &  Ic  & 0.2285 & 11   &  \\  
  \hline 
 SDSS 2004hx   &  \IIP\  & 0.0375 & 12   &  \\  
 SDSS 2004ib   &  Ib  & 0.0555 & 13   &  \\  
 SDSS 2005hm   &  Ib  & 0.0339 & 14   &  \\  
 SDSS 2005gi   &  \IIP\  & 0.0494 & 15   & \checkmark \\  
 SDSS 004012\footnotemark[3]  &  Ic  & 0.0246 & 16   &  \\  
 SDSS 2006ez   &  \IIn\  & 0.0876 & 17   &  \\  
 SDSS 2006fo   &  Ic  & 0.0199 & 18   &  \\  
 SDSS 2006gq   &  \IIP\  & 0.0688 & 19   & \checkmark \\  
 SDSS 2006ix   &  \IIn\  & 0.0745 & 20   &  \\  
 SDSS 2006kn   &  \IIP\  & 0.1193 & 21   & \checkmark \\  
 SDSS 014475\footnotemark[3]  &  Ic  & 0.1425 & 22   &  \\  
 SDSS 2006jo   &  Ib  & 0.0757 & 23   &  \\  
 SDSS 2006jl   &  \IIP\  & 0.0546 & 24   & \checkmark \\  
 SDSS 2006iw   &  \IIP\  & 0.0295 & 25   & \checkmark \\  
 SDSS 2006kv   &  \IIP\  & 0.0608 & 26   & \checkmark \\  
 SDSS 2006ns   &  \IIP\  & 0.1189 & 27   & \checkmark \\  
 SDSS 2006lc   &  Ic  & 0.0150 & 28   &  \\  
 SDSS 2007ms   &  Ic  & 0.0384 & 29   &  \\  
 SDSS 2007iz   &  \IIP\  & 0.2525 & 30   &  \\  
 SDSS 2007nr   &  \IIP\  & 0.1433 & 31   & \checkmark \\  
 SDSS 2007kw   &  \IIP\  & 0.0672 & 32   & \checkmark \\  
 SDSS 2007ky   &  \IIP\  & 0.0725 & 33   & \checkmark \\  
 SDSS 2007lj   &  \IIP\  & 0.0489 & 34   & \checkmark \\  
 SDSS 2007lb   &  \IIP\  & 0.0326 & 35   & \checkmark \\  
 SDSS 2007ll   &  \IIP\  & 0.0801 & 36   &  \\  
 SDSS 2007nw   &  \IIP\  & 0.0562 & 37   & \checkmark \\  
 SDSS 2007ld   &  \IIP\  & 0.0260 & 38   & \checkmark \\  
 SDSS 2007md   &  \IIP\  & 0.0535 & 39   & \checkmark \\  
 SDSS 2007lz   &  \IIP\  & 0.0928 & 40   & \checkmark \\  
 SDSS 2007lx   &  \IIP\  & 0.0556 & 41   & \checkmark \\  
 SDSS 2007og   &  \IIP\  & 0.1995 & 42   &  \\  
 SDSS 2007ny   &  \IIP\  & 0.1452 & 43   & \checkmark \\  
 SDSS 2007nv   &  \IIP\  & 0.1427 & 44   & \checkmark \\  
 SDSS 2007nc   &  Ib  & 0.0856 & 45   &  \\  

\tableline  
\end{tabular}
\end{center}
\footnotetext[1]{Non-Ia index used in the \SNPCC.}
\footnotetext[2]{\checkmark means the \IIP\ light curve has been
     publicly available in  \citet{SDSS_IIP_2010} since 2010 Jan 1.}
\footnotetext[3]{Internal SDSS index.}
  \label{tb:nonIa_list}
\end{table}

 \subsection{SN Rates and Template Weights}
 \label{subsec:sim_rates}

Following \citet{Dilday2008}, the SN~Ia volumetric rate ($r_V$) 
was parametrized as $r_v = \alpha(1+z)^{\beta}$ with
$\alpha_{\rm Ia} = 2.6\times 10^{-5}$~
${\rm Mpc}^{-3}\,h_{70}^3\,{\rm year}^{-1}$,
$\beta_{\rm Ia} = 1.5$, and 
$h_{70} = H_0/(70\,{\rm km}\,s^{-1}\,{\rm Mpc}^{-1})$  
where $H_0$ is the present value of the Hubble parameter.
Integrating out to a redshift of $z = 1.1$, 
the total number of generated SN~Ia for the DES survey is 
$\sim \NGENIa$, and the number written for the \SNPCC\ 
(i.e., passing the loose cuts in \S\ref{subsec:sim_overview}) 
is $\sim \NCUTIa$.

For the \nonIa\ rate, we assumed that the redshift dependence
has the same general form as for the SNe~Ia.
The exponent term $\beta_{\rm nonIa} = 3.6$ was taken to
match that of the star formation rate. 
To estimate $\alpha_{\rm nonIa}$ we use the result
of \citet{Bazin2009} which reports an observed non-Ia/Ia 
rate ratio of $4.5 \pm 1.0$ for $z<0.4$.  
We then calculate $\alpha_{\rm nonIa} = 6.8\times 10^{-5}$ 
such that the  {\nonIa}/Ia rate ratio
matches the observed ratio. 
Since the \nonIa\ rate has a much larger \unc\ at 
redshifts above 0.4, and to 
increase the sample of misclassified \nonIa,
the \nonIa\ rate was arbitrarily increased by a 
factor of 1.3 at all redshifts.
Integrating out to a redshift of $z = 1.1$, 
the total number of generated \nonIa\ for the DES survey is 
$\sim \NGENnonIa$, and the number written out for the \SNPCC\
is $\sim \NCUTnonIa$.

The generated non-Ia/Ia ratio over all redshifts is 12.
After applying the loose selection requirements for the 
\SNPCC\ sample (\S\ref{subsec:sim_overview}),
this ratio drops to 2.4. We have likely overestimated
the \nonIa\ contribution, but this overestimate was 
intentional in order to increase the statistics of 
\nonIa\ SNe that are misidentified as SN~Ia.

The breakdown of the \nonIa\ into subtypes 
(\Ibc,\IIP,\IIL, and \IIn) is taken
from  \citet{Smartt2009}, and the subtype fractions
are shown in Table~\ref{tb:nonIa_subtypes} along
with the number of templates used to represent
each subtype. Within a subtype class, each \nonIa\ template
is given equal weight in the generation of simulated samples.
For each subtype a composite Nugent template is included,
and is given the same generation weight as each
template based on an observed light curve.

\begin{table}
\caption{Non-Ia Subtype Fractions and Template Statistics}
\begin{center}
\begin{tabular}{lccc}
\tableline\tableline
             &            & No. of     & No. of      \\
  Non-Ia     &            & measured   & composite   \\
  subtype    & Fraction   & templates  & templates   \\
\tableline
   \Ibc\     & \ProbIbc\  & \NumIbc\   &  1      \\
   \IIP\     & \ProbIIP\  & \NumIIP\   &  1      \\
   \IIL\     & \ProbIIL\  & \NumIIL\   &  1      \\
   \IIn\     & \ProbIIn\  & \NumIIn\   &  1      \\
\tableline  
\end{tabular}
\end{center}
  \label{tb:nonIa_subtypes}
\end{table}

 \subsection{Spectroscopic Subset}
 \label{subsec:sim_spec_subset}

\begin{figure}
\centering
\epsscale{1.1}  
\plotone{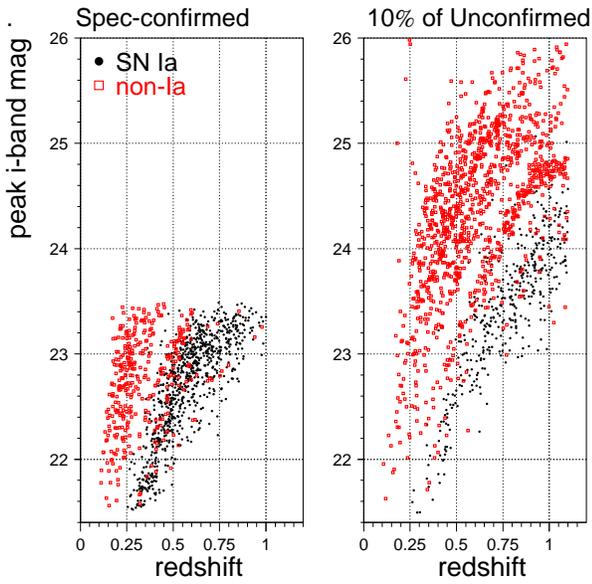}
  \caption{
    For SNe with $r > 21.5$ at peak brightness,
    peak $i$-band magnitude vs. redshift for
    the \specy\ confirmed subset (left) and for 10\% of the 
    unconfirmed sample (right). The SNe~Ia are shown by
    filled black circles; the \nonIa\ SNe by open red squares.
    The dashed grid lines are shown to guide the eye.
  }
  \label{fig:imag}
\end{figure}

To allow participants to train their
classification algorithms,
a \specy\ confirmed  training subset was provided.
This subset was based on \obss\ from a 4~m class telescope 
with a limiting $r$-band magnitude of $\rspeclimit$, 
and on {\obss} from an 8~m class telescope
with a limiting $i$-band magnitude of $\ispeclimit$.
Using this \spec\ selection resulted in a 
subset of $\NSPEC$ objects, or about 7\% of the total number
of objects in the \SNPCC.
This training sample is not a random subset
and is, in fact, a highly biased subset,
as shown in Fig.~\ref{fig:imag};
the true SN~Ia fraction for the confirmed SNe is 70\%, 
compared with only 26\% for the unconfirmed SNe.
While a truly random subset would be ideal for
training classification algorithms,
limited  \spec\ resources in future surveys 
are much more likely to obtain a biased sample
unless there is sufficient motivation to modify
the \spec\ targeting strategy.

If each SN spectrum were taken exactly at
the epoch of peak brightness ($t_0$), then the
\eff\ for obtaining a spectrum adequate for classification
would depend only on the peak magnitude.
However, a spectrum is typically  taken slightly before
or after $t_0$, when the SN is slightly dimmer than
at peak brightness;
therefore we have parametrized the
efficiency for obtaining a spectrum ($\effspec$) to be
\begin{equation}
  \effspec = \epsilon_0(1-x^{\ell})~,~~~ 
   x \equiv \frac{\magpeak - \magmin}{\maglim - \magmin}~,
\end{equation}
where the parameters $\ell$, $\magmin$ and $\maglim$ 
are given in Table~\ref{tb:effspec}
for the $r$ and $i$ filters,
and $\magpeak$ is the SN magnitude at $t_0$.
The coefficient $\epsilon_0 = 0.4$ for type Ia
and 0.3 for \nonIa; 
this difference in the $\epsilon_0$  values was due to
an error in the simulation (\S\ref{subsec:sim_bugs}).
The efficiency function is nearly flat for bright SNe
and then decreases rapidly to zero at the limiting magnitude.
A simulated SN is \specy\ identified if 
$21.5 < \magpeak^i< \ispeclimit$ and a randomly generated number 
(0--1) is less than $\effspec^i$, or if the analogous criterion
is satisfied for the $r$ band.
Since the $\effspec$ parametrization is an educated guess,
future simulations should use a more refined parametrization 
based on the range of epochs in which
\spec\ \obss\ are expected to be obtained.

\begin{table}
\caption{Efficiency Parameters for Spectroscopic Observations}
\begin{center}
\begin{tabular}{l | lll}
  Filter  &  $\ell$  &  $\magmin$   &  $\maglim$ \\   
\tableline\tableline
   $r$    &    5     &  16.0  & $\rspeclimit$  \\
   $i$    &    6     &  21.5  & $\ispeclimit$   \\
\tableline  
\end{tabular}
\end{center}
  \label{tb:effspec}
\end{table}

 \bigskip
 \subsection{Bugs}
 \label{subsec:sim_bugs}

Here, we begin with the bugs that were 
identified and fixed
before the \SNPCC\ deadline for submissions;
we then report bugs that were present during
the \SNPCC\ and fixed after the submission deadline.
The identification of bugs by the participants resulted
in three updates during the \SNPCC.
For each update, only a small ($\sim 1$\%) fraction
of the sample was modified, although the last update
resulted in a 10\% reduction in the sample size.
A summary of bugs is shown in Table~\ref{tb:bugs}.

The first bug resulted in a small fraction of the SNe~Ia having
late-time fluxes that were much larger than the flux at the 
nominal epoch of peak brightness.
This bug was induced by a poorly constrained 
quadratic term for the shape parameter correction
in the \mlcsU\ model,\footnote{See the $Q$ parameter in \citet{JRK07}.}
and it only affected fast-declining 
SNe~Ia at epochs well past peak brightness.
This artifact was removed by introducing a damping function
for the quadratic term.

The second bug resulted in a small fraction of the \nonIa\ SNe
being much brighter than the SNe~Ia. This bug was caused by
using an untruncated Gaussian distribution to select 
random magnitudes for the small fraction of \nonIa\ 
based on the Nugent SED templates.
This bug was fixed by requiring the random numbers to lie
within $\pm 2\sigma$ of the mean.

The next issue involved an ambiguous redshift for
SDSS SN 2004hx. The original redshift used to 
make the SED template was based on the host galaxy
($z_{\rm host} = 0.0382$)
and led to an exceptionally bright type II SN.
However, the preliminary redshift from the SN spectrum
is $z_{\rm SN} \simeq 0.014$,
suggesting a type II SN with normal brightness.
During the \SNPCC\ we changed this SED template to use 
the normal SN brightness and left the redshift ambiguity 
to be resolved in a future analysis.

The remaining four bugs described below were not corrected
until after the \SNPCC. The first unfixed bug is related
to the rest-frame wavelength ranges covered by the SN
models.
While the \nonIa\ models are defined for all rest-frame
wavelength ranges, the valid wavelength range of both 
SN~Ia models was restricted to be above 2500~{\AA}.
This wavelength restriction resulted in undefined 
$g$-band model magnitudes for SNe~Ia at $z > 0.8$.
To warn users about observations with undefined model
magnitudes, the \SNANA\ simulation treats undefined 
model values by writing the flux as $-9 \pm -9$.
This feature of the simulation was not noticed during 
the preparation of the \SNPCC, and therefore high-redshift 
SNe can be identified by simply inspecting the $g$-band 
flux value.
For \SNPCC\ participants who included these invalid $g$-band 
fluxes as if they were valid measurements, 
the absolute value of the \unc\ is a few times larger 
than the sky noise. Therefore, by accidental
good luck this invalid value is consistent with the 
correct value based on the sky noise and the very 
small SN~Ia flux expected in the far-ultraviolet region.

There were significantly more SN~Ia generated by the \SALTII\
model than by the \mlcsU\ model. The primary reason is that
we mistakenly used symmetric color and stretch distributions
for \SALTII, while using the measured asymmetric distributions
for \mlcsU. The missing non-Gaussian tails in the \SALTII\
distributions resulted in an SN~Ia sample that was $\sim 0.2$~mag
too bright on average. This issue is discussed further in
\S\ref{sec:sim_update}.

This next bug is by far the most embarrassing.
Each \nonIa\ SED template is too dim by a factor of 
${1+\zobs}$, where $\zobs$ is the observed redshift of 
the \nonIa\ SN used to construct the template;
note that $\zobs$ is {\it not} the simulated redshift.
Thus for a \nonIa\ template constructed from an SN at $\zobs=0.1$,
all simulated SN based on this template were 10\% too dim.
Figure~\ref{fig:smooth_non1a} shows that some of the 
most commonly misidentified \nonIa\ light curves in the
\SNPCC\ were based on SDSS SNe with $0.1 < \zobs < 0.25$, 
and therefore these simulated \nonIa\ SNe were 10--25\% too dim.
The combination of SNe~Ia that are too bright (previous bug) 
and \nonIa\ SNe that are too dim may have made the photometric 
challenge somewhat easier  for some methods.

To improve analysis efficiency, the \SNANA\ simulation was
originally designed to exclude pre-explosion epochs.
Although pre-explosion epochs should have been included
in the \SNPCC\ sample, we did not notice the missing epochs
until one of the participants acknowledged using this feature 
to estimate the time of peak brightness.

As described in \S\ref{subsec:sim_spec_subset}, 
the \specy\ confirmed fraction was 
different for the SN types:
for SN passing the \spec\ magnitude limits,
the type Ia SNe were confirmed 33\% more often than the \nonIa.
The last known bug is that there is a trivial way to
identify each SN type without any knowledge of SN science.
After all submissions had been received,
an ``SN Cheater Challenge'' was offered on 2010 June 2;
it was solved 16 hours later by Sako
(see Table~\ref{tb:participants}),
but so far nobody else has solved it.

\begin{table}
\caption{
	Summary of Bugs in the \SNPCC\ Simulation.
	}
\begin{center}
\begin{tabular}{ll}
\tableline\tableline
 Date of        &                      \\
 bug fix        & Description of bug   \\
\tableline  
  Mar 14, 2010       & Enormous fluxes for late-time \\
                     & (fast-declining) SNe~Ia \\
                     & generated with \mlcsU\  \\  \hline
  Mar 24, 2010      & Extremely bright \nonIa\ from       \\
                    & untruncated Gaussian smearing \\
                    & in Nugent template mags       \\ \hline
  Apr 13, 2010      & Ambiguous redshift for 2004hx \\ \hline
  After \SNPCC\      & $g$-flux and error are $-9$ for \\
                    & SNe~Ia with $z>0.8$             \\ \hline
  After \SNPCC\      & Average \SALTII\ SN~Ia is 0.2~mag too \\
                    &  bright due to missing tails  \\ \hline
  After \SNPCC\      & Each \nonIa\ SED template is    \\
                    & too dim by a factor of $1+\zobs$    \\ \hline
  After \SNPCC\      & no pre-explosion epochs \\ \hline 
  After \SNPCC\      & \Spec\ fractions were  \\  
                    & different for Ia and \nonIa\   \\ \hline
  Not fixed         & Trivial to cheat on entire \\
                    &  \SNPCC\ sample            \\
\tableline  
\end{tabular}
\end{center}
  \label{tb:bugs}
\end{table}

 \section{Taking the SN Classifier Challenge}
 \label{sec:take_challenge}

As described in \S\ref{sec:sim},
two independent challenges were generated:
one with a host-galaxy \photoz\ for each SN
and another without any redshift information.
In addition to these challenges based on the entire light curve, 
there was also an early-epoch challenge
motivated by the need to prioritize SNe for \spec\ 
follow-up observations;
this challenge was based on the first six photometric
observations (in any filter) 
with ${\rm S/N} > 4$. 
Participants attempted the full light-curve challenges 
with and without redshift information, but none of the
participants attempted the early-epoch challenge,
due to time limitations and the increased interest on
the full light curve challenge that will eventually
impact the cosmology analyses.

The simulated light curves are available at the
\SNPCC\ Web site.\footnote{\wwwdownload}
Details on how to analyze the simulated sample
are given in {\S 3} of  the \SNPCC\ release note.
To fully optimize classification algorithms during the challenge, 
several participants wanted to know the exact value of the 
false-tag weight (\S\ref{sec:eval}) used to determine 
the figure of merit.
On 2010 April 27 we therefore publicly announced that 
$\wIafalse = \wIafalseVALUE$;
while this information clearly helped some participants
optimize results for the confirmed subset, it is not clear
if the information improved results for the unconfirmed sample.

A total of \NGROUP\ groups (or individuals) sent
\NSUBMIT\  submissions to be evaluated.
Among the submissions, \NSUBMIThostz\ are based
on the \SNPCCHOSTZ, while the remaining \NSUBMITnohostz\ 
are based on the \SNPCCnoHOSTZ.
Photo-$z$ estimates were given by four participants
in the \SNPCCHOSTZ 
and by three participants in the \SNPCCnoHOSTZ.

Table~\ref{tb:participants} shows the list of groups and 
participants, indicates which challenge(s) were taken, 
and indicates if SN \photoz\ estimates were given. 
The average processing time is also given for each method, 
and these times vary from 1~s to $>200$~s per SN
using similar processors.
A brief description for each method is given in
Appendix~\ref{app:methods}.

Among the participants, four general strategies were used to
classify SNe. The first and simplest strategy was to fit 
each light curve to an SN~Ia model 
and use the 
``duck test'' philosophy:
if it looks like a duck (i.e., an SN~Ia) and quacks like a duck, 
then it is a duck.
Selection cuts, mainly on the minimum $\chi^2$,
were used to determine which SNe are type Ia,
and there was no attempt to classify a subtype for \nonIa.
This strategy was used by 
Gonzalez, Portsmouth-$\chi^2$ and SNANA cuts.

The second strategy compares each light curve 
against both SN~Ia and \nonIa\ templates, 
and uses the Bayesian probabilities to determine the 
most likely SN type.
Poz2007 used the simplest Bayesian implementation, 
with a single Ia and \nonIa\ template. 
Belov \& Glazov and Sako used SN~Ia templates that depend
on stretch and extinction, and they also used several 
\nonIa\ templates. 
Sako included 8 \nonIa\ templates from the \SDSS,
although there was no coordination between his 
template development for classification and the 
development of templates for the \SNPCC.
Rodney used a variant of this technique by accounting for the 
fact that templates from observed SNe do not form a complete set.
MGU+DU used another variation by using slopes (mag/day) at
four different epochs and comparing with slopes expected
for type Ia and \nonIa\ SNe.

The third strategy used  \specy\ confirmed SNe~Ia
to parametrize a Hubble diagram, and then identified
SN~Ia as those SNe that lie near the expected Hubble diagram.
Portsmouth-Hub used a high-order polynomial to define
the Hubble diagram while JEDI-Hub used the 
kernel density estimation technique.

In the last strategy (InCA and JEDI-KDE) each light curve 
was fit with a parametric function such as a spline, and 
the fitted parameters were used for
statistical inferences.
Light-curve fitting parameters such as 
stretch and color were not explicitly used.

\begin{table*}
\caption{
	List of Participants in the \SNPCC.
  } 
\begin{center}
\begin{tabular}{ll c l c l }
\tableline\tableline
             &                               
   & Classified  & SN         &              \\
     Participants      & Abbreviation\footnotemark[1]  
   & +Z\footnotemark[2]/noZ\footnotemark[3]    
   & $z_{\rm ph}$\footnotemark[4]
   & CPU\footnotemark[5]
   &  Description (strategy class\footnotemark[6]) \\
\hline\hline
P. Belov and S. Glazov 
            & Belov \& Glazov   & yes/no    & no   & 90
            & light curve $\chi^2$ test against Nugent templates (2) \\  \hline
S. Gonzalez & Gonzalez       & yes/yes   & no  & 120
            & cuts on SiFTO fit $\chi^2$ and fit parameters (1) \\  \hline
J. Richards, Homrighausen, 
              & InCA\footnotemark[7]  & no/yes  & no  & 1
              & Spline fit \& nonlinear dimensionality  \\
C. Schafer, P. Freeman   & & & & & reduction (4)  \\  \hline
J. Newling, M. Varuguese, 
      & JEDI-KDE & yes/yes  & no  & 10
      & Kernel Density Evaluation with 21 params (4) \\ 
B. Bassett, R. Hlozek, 
      & JEDI Boost  & yes/yes  & no  & 10
      & Boosted decision trees (4) \\ 
D. Parkinson, M. Smith, 
      & JEDI-Hubble  & yes/no  & no  & 10
      & Hubble diagram KDE (3) \\ 
H. Campbell, M. Hilton, 
      & JEDI Combo  & yes/no  & no  & 10
      & Boosted decision trees + Hubble KDE (3+4) \\ 
H. Lampeitl, M. Kunz,     
          & & & & \\ 
P. Patel ~~(JEDI group\footnotemark[8])
          & & & & \\  \hline
S. Philip, V. Bhatnagar, 
            & MGU+DU-1\footnotemark[9]            
            & no/yes & no  & $<1$
            & light curve slopes \& Neural Network (2) \\
A. Singhal, A. Rai,   
            & MGU+DU-2 
            & no/yes & no   & $<1$
            & light curve slopes \& Random Forests (2)  \\ 
A. Mahabal, K. Indulekha  
            & & & & \\  \hline
H. Campbell, B. Nichol, & Portsmouth $\chi^2$ & yes/no & no  & 1
           & SALT2--$\chi^2_r$ \& False Discovery Rate Statistic  (1) \\
H. Lampietl, M .Smith   & Portsmouth-Hubble & yes/no & no  & 1
           & Deviation from parametrized Hubble diagram (3)  \\  \hline
D. Poznanski 
            & Poz2007 RAW     & yes/no    & yes   & 2
            & SN Automated Bayesian Classifier (SN--ABC)  (2) \\ 
            & Poz2007 OPT     & yes/no    & yes   & 2  
            & SN--ABC with cuts to optimize $\FoMIa$ (2).   \\   \hline
S. Rodney   & Rodney         & yes/yes   & yes  & 230
            & SN Ontology with Fuzzy Templates  (2) \\ \hline
M. Sako 
            & Sako           & yes/yes   & yes   & 120
            & $\chi^2$ test against grid of Ia/II/Ibc templates  (2) \\  \hline
S. Kuhlmann, R. Kessler 
            & SNANA cuts       & yes/yes   & yes   & 2
            & Cut on {\mlcs} fit probability, S/N \& sampling (1) \\   
\tableline  
\end{tabular}
\end{center}
\footnotetext[1]{Groups are listed alphabetically by abbreviation.}
\footnotetext[2]{Classifications included for \SNPCCHOSTZ.}
\footnotetext[3]{Classifications included for \SNPCCnoHOSTZ.}
\footnotetext[4]{\photoz\ estimates included.}
\footnotetext[5]{Average processing time per SN (seconds) 
                 using similar 2-3 GHz cores.}
\footnotetext[6]{From \S\ref{sec:take_challenge}, strategy classes are
             1) selection cuts, 2) Bayesian probabilities,
             3) Hubble-diagram parametrization and 4) statistical inference.
        }
\footnotetext[7]{International Computational Astrophysics Group: \wwwINCA}
\footnotetext[8]{Joint Exchange and Development Initiative: \wwwJEDI}
\footnotetext[9]{MGU=Mahatma Gandhi University, DU=Delhi University.}
  \label{tb:participants}
\end{table*} 

 \section{Evaluating the {\SNPCC} }
 \label{sec:eval}

Ideally we would like to assign a single number,
or figure of merit (FoM), for each \SNPCC\ submission.
We begin the discussion by considering a measurement  of 
the SN~Ia rate based on \phot\ \id.
After selection requirements have been applied,
let $\NIatrue$ be the number of correctly typed SNe~Ia,
and $\NIafalse$ be the number of non-Ia that are 
incorrectly typed as an SN~Ia. A simple classification FoM 
is the square of the S/N divided
by the total number of SNe~Ia ($\NIaTOT$) before selection cuts,
\begin{eqnarray}
    \FoMIa &  \equiv & 
     \frac{1}{\NIaTOT} \times 
     \frac{({\NIatrue})^2}{\NIatrue + \wIafalse\NIafalse}  
          \nonumber  \\
       &    &  \nonumber \\
       &  = & \frac{\NIatrue}{\NIaTOT} \times
              \frac{\NIatrue}{\NIatrue + \wIafalse\NIafalse} ~,
          \nonumber  \\
       &  = & \effIa \times \PurityIa~,
   \label{eq:FoM}
\end{eqnarray}
where $\effIa = \NIatrue/\NIaTOT$ is the SN~Ia efficiency that 
includes both selection and classification requirements,
$\PurityIa$ is the pseudopurity, and
$\wIafalse$ is the false-tag weight (penalty factor).
Since $\NIaTOT$ is a constant that is independent of the analysis, 
we have divided out this term so that $0 \le \FoMIa \le 1$, 
with $\FoMIa = 1$ corresponding to the theoretically optimal analysis.

When $\wIafalse=1$, 
the denominator in $\PurityIa$ comes from the Poisson noise term
in the S/N, and $\PurityIa$  can be interpreted as the 
traditional purity factor defined as the fraction of 
classified Ia that really are SNe~Ia.
In the ideal case where the mean of $\NIafalse$ is 
perfectly determined,\footnote{
The mean $\NIafalse$ value is the
average over many independent measurements.
} 
the naive Poisson \unc\ is the only contribution to the 
noise term and therefore $\wIafalse = 1$.
In practice, however, \uncs\ in determining the false-tag rate 
lead to $\wIafalse > 1$. For example, suppose that the estimate 
of $\NIafalse$ is scaled from a \specy\ confirmed subset 
containing a fraction ($\effspec$) of the total number of SNe;
in this case, the Poisson noise term is defined by setting 
$\wIafalse = 1 + \effspec^{-1}$, and $\wIafalse \gg 1$ if the 
\spec\ subset is small.

When using SN~Ia for cosmological applications,
it may be possible to reduce $\wIafalse$ using other methods 
to determine $\NIafalse$, such as fitting the tails in the 
distance-modulus residuals. 
A proper determination of  $\wIafalse$ is beyond the scope of this 
classification challenge, and we have therefore arbitrarily set
$\wIafalse = \wIafalseVALUE$. While this value is well below
$1/\effspec \sim 15$ based on using the \specy\ confirmed subset,
$\wIafalse$ is notably larger than unity and therefore 
penalizes incorrect classifications more than rejected SNe.

 \bigskip
 \section{Results}
 \label{sec:results}

Here, we give a relatively brief overview of the main results 
and comparisons. 
Ideally, we would fully understand the strengths and weaknesses 
for each entry, but this level of detail is deferred to 
future analyses from individual participants.
Also, since the results presented here are simply a starting point
for these studies, a detailed postchallenge analysis could soon 
become obsolete as the algorithms are improved.
Finally, the most important goal here is not to identify the best
method now, but to motivate improvements and then identify the
best method appropriate to each SN survey.

\begin{figure*}
\centering
\epsscale{0.8}
\plotone{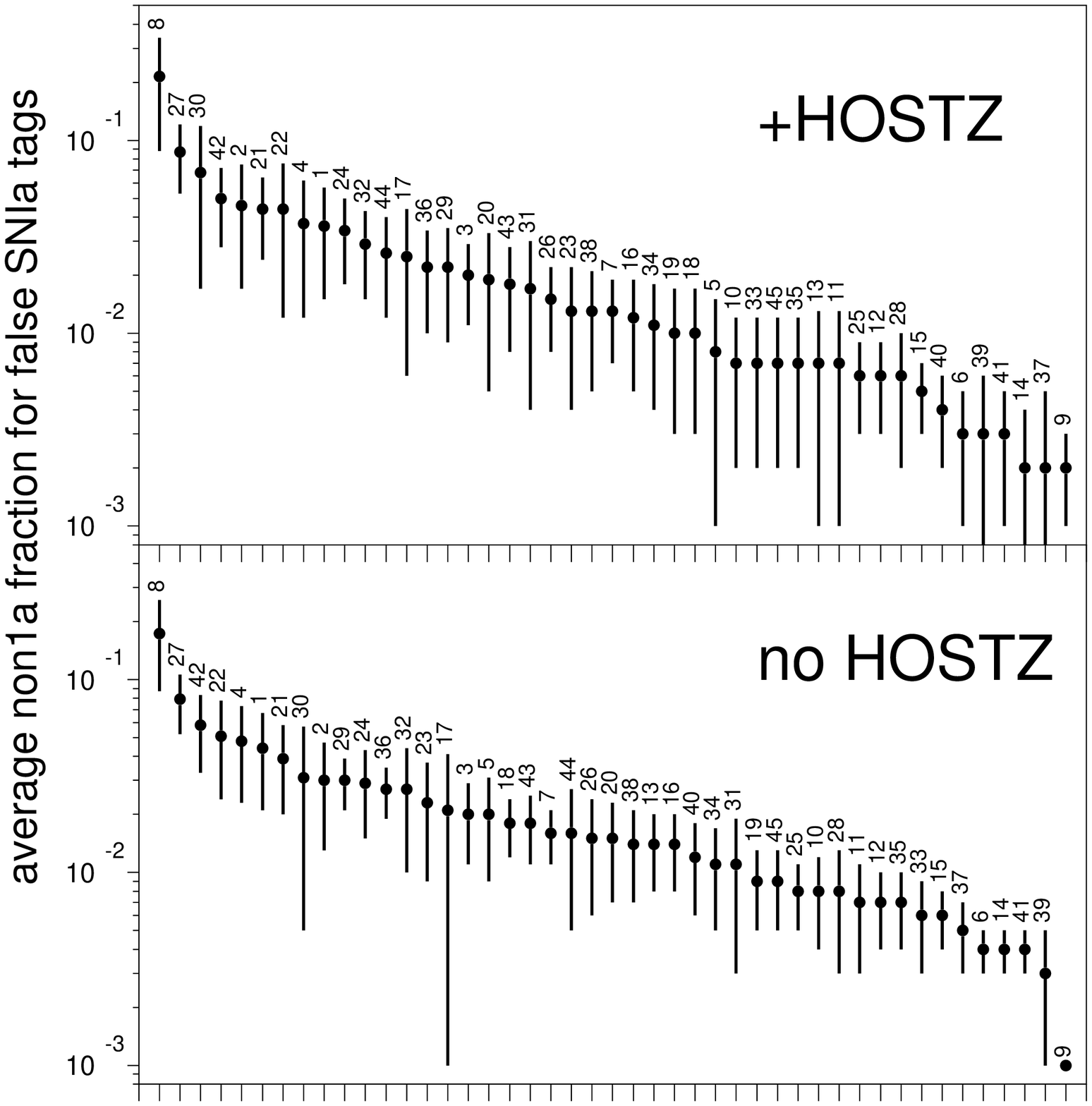}
  \caption{
	Among all generated \nonIa\ SNe that were falsely 
	tagged as an SN~Ia,
	the average fraction of each \nonIa\ template is shown.
        Error bars reflect the rms dispersion among the submissions.	
	The \SNPCC\ index shown above each data point is translated
	  in Table~\ref{tb:nonIa_subtypes}.
      }
  \label{fig:non1afrac}
\end{figure*}

We begin by showing the \nonIa\ SNe
that were misidentified as SNe~Ia.
For each challenge entry we have computed the fraction of 
false SN~Ia tags corresponding to each \nonIa\ SED template: 
the sum of these fractions equals one for each entry.
Fig.~\ref{fig:non1afrac} shows the false-tag fractions
averaged over all entries, 
and they are sorted from largest to smallest.
For both challenges (with and without host-galaxy {\photoz}),
the most frequently misidentified \nonIa\ 
is based on SN 2006ep
(\SNPCC\ index $=8$; see Table~\ref{tb:nonIa_subtypes}),
a \specy\ confirmed SN~Ic with a rest-frame $g$-band  
peak magnitude of $-19.1$~mag.
While the generated fraction for each \Ibc\ SED template is 
1.7\% of the total,
simulated \nonIa\ SNe based on 2006ep account for 
$\sim 20$\%  of all misidentified SN~Ia.
The second most frequently misidentified \nonIa\ template,
accounting for 8\% of all falsely tagged SN~Ia,
is based on SN 2006ns (\SNPCC\ index $=27$),
a \specy\ confirmed type II-P SN with a $g$-band peak magnitude
of $-18.3$~mag.

\begin{figure*}
\centering
\epsscale{1.1}
\plottwo{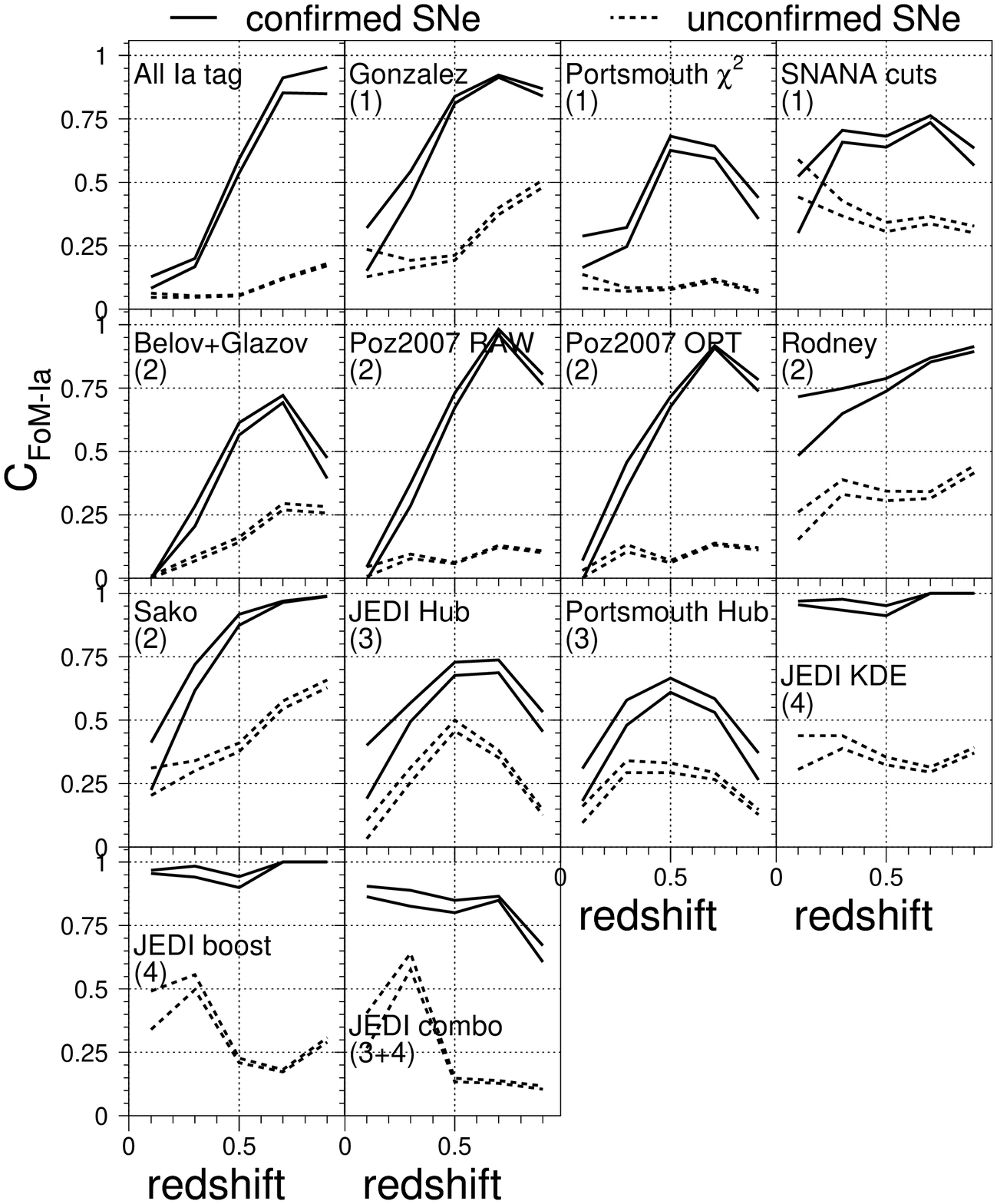}{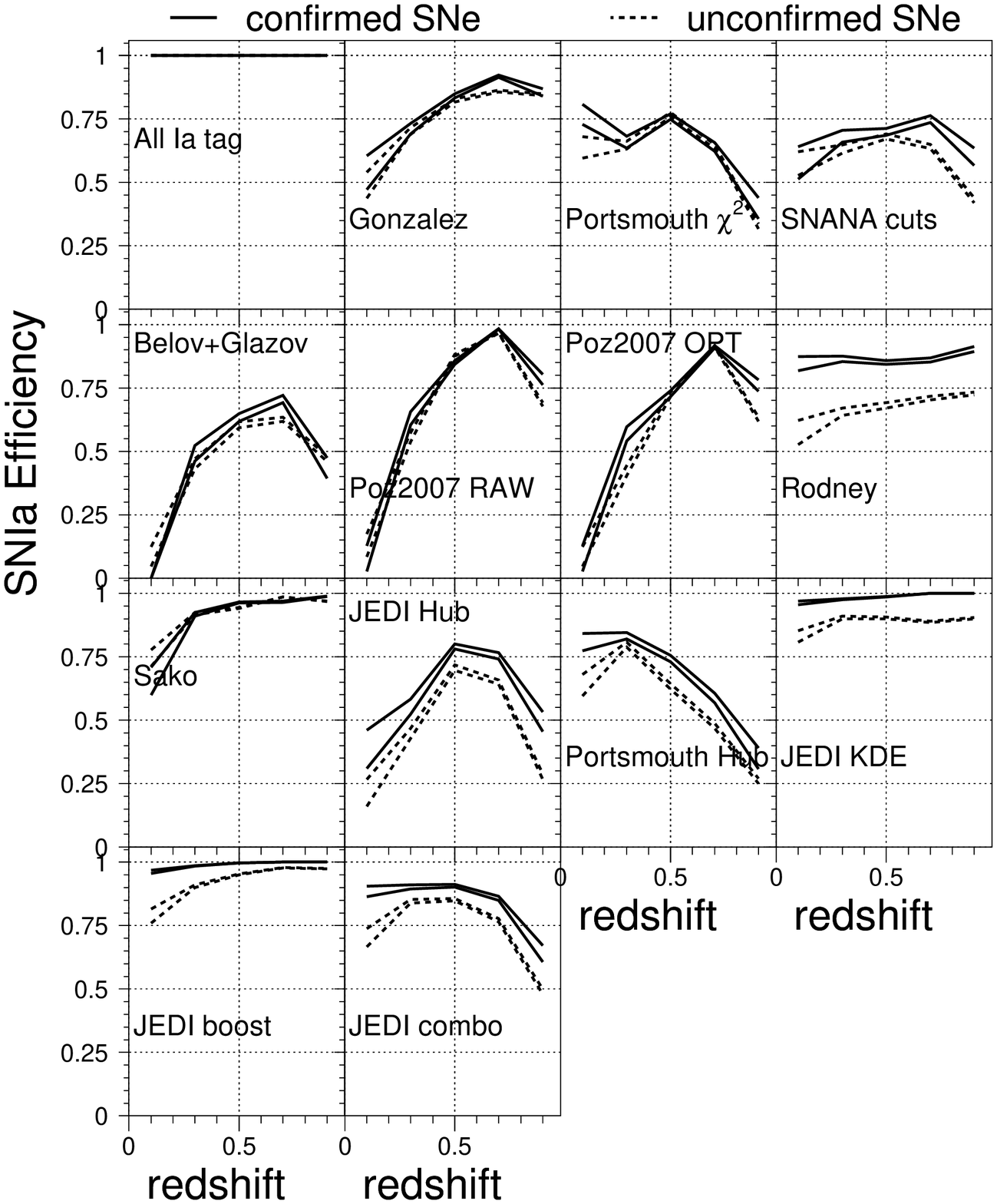}  
\plottwo{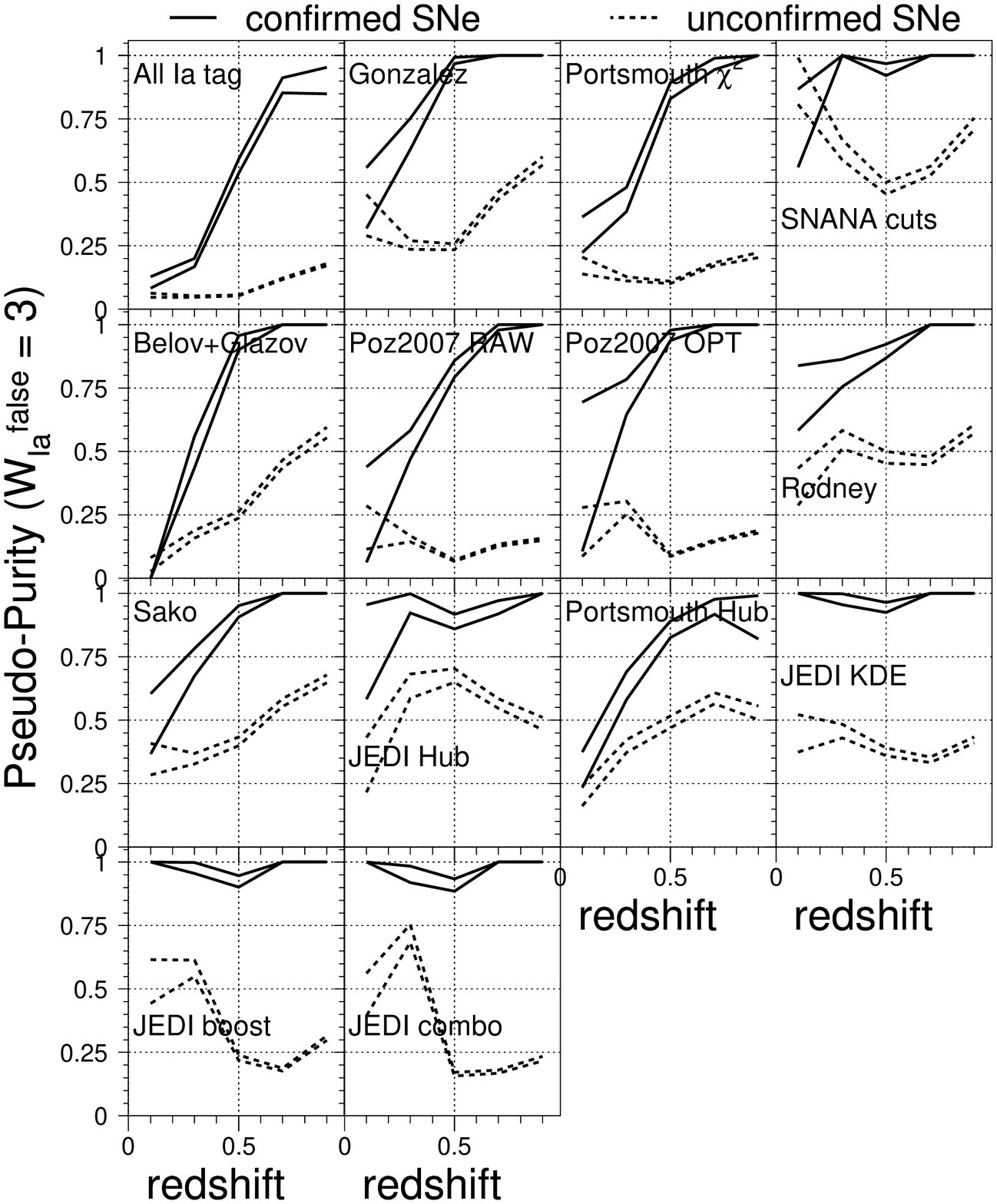}{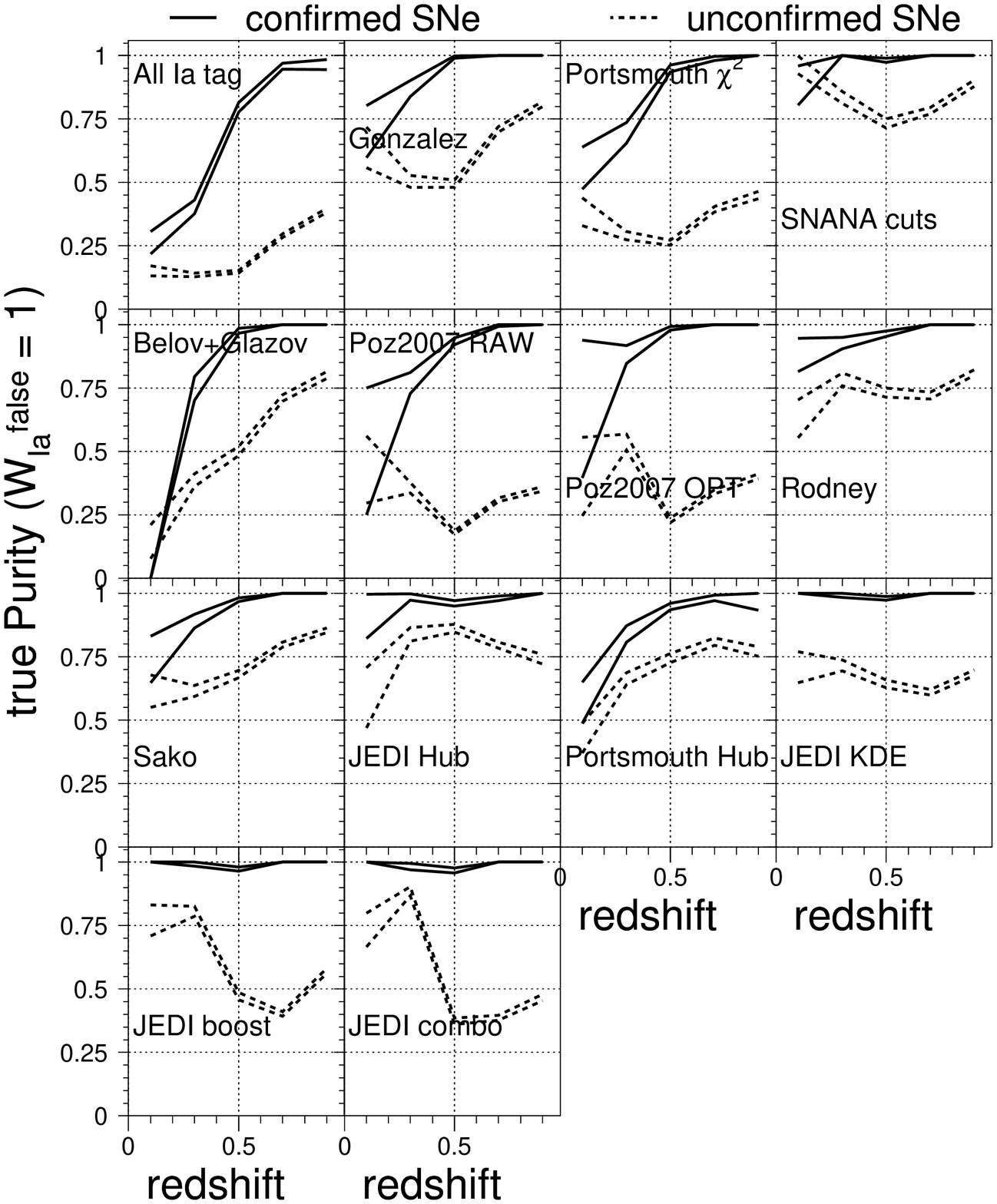}  
  \caption{
    For each participant in the \SNPCCHOSTZ,
    results vs. redshift are shown for $\FoMIa$, $\effspec$,
    pseudopurity ($\PurityIa$), and the true purity
    ($\PurityIa$ with $\wIafalse=1$).
    {\curvedefSNIa}
      }
  \label{fig:hostz_resultsIa}
\end{figure*}

\begin{figure*}
\centering
\epsscale{1.1}
\plottwo{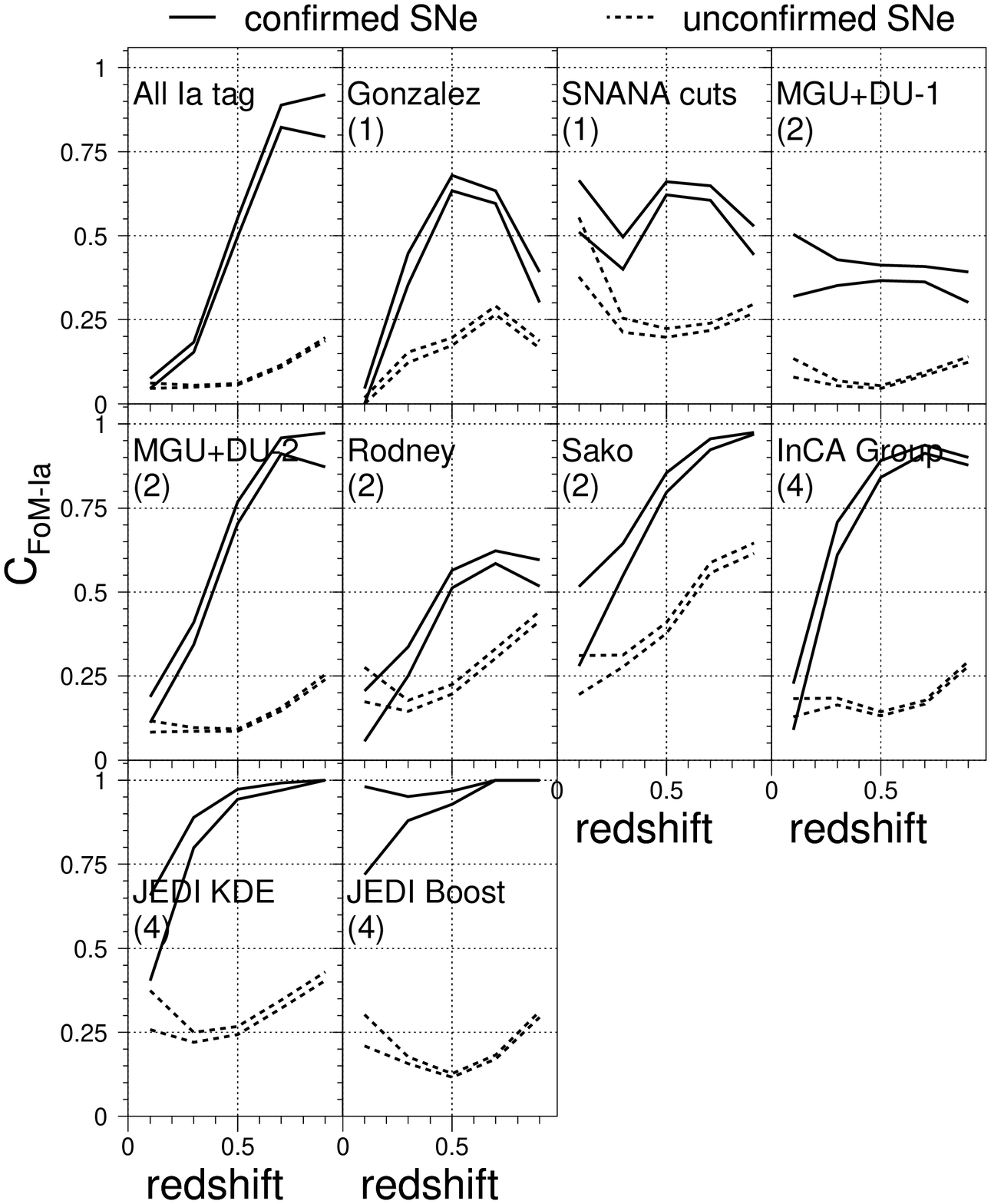}{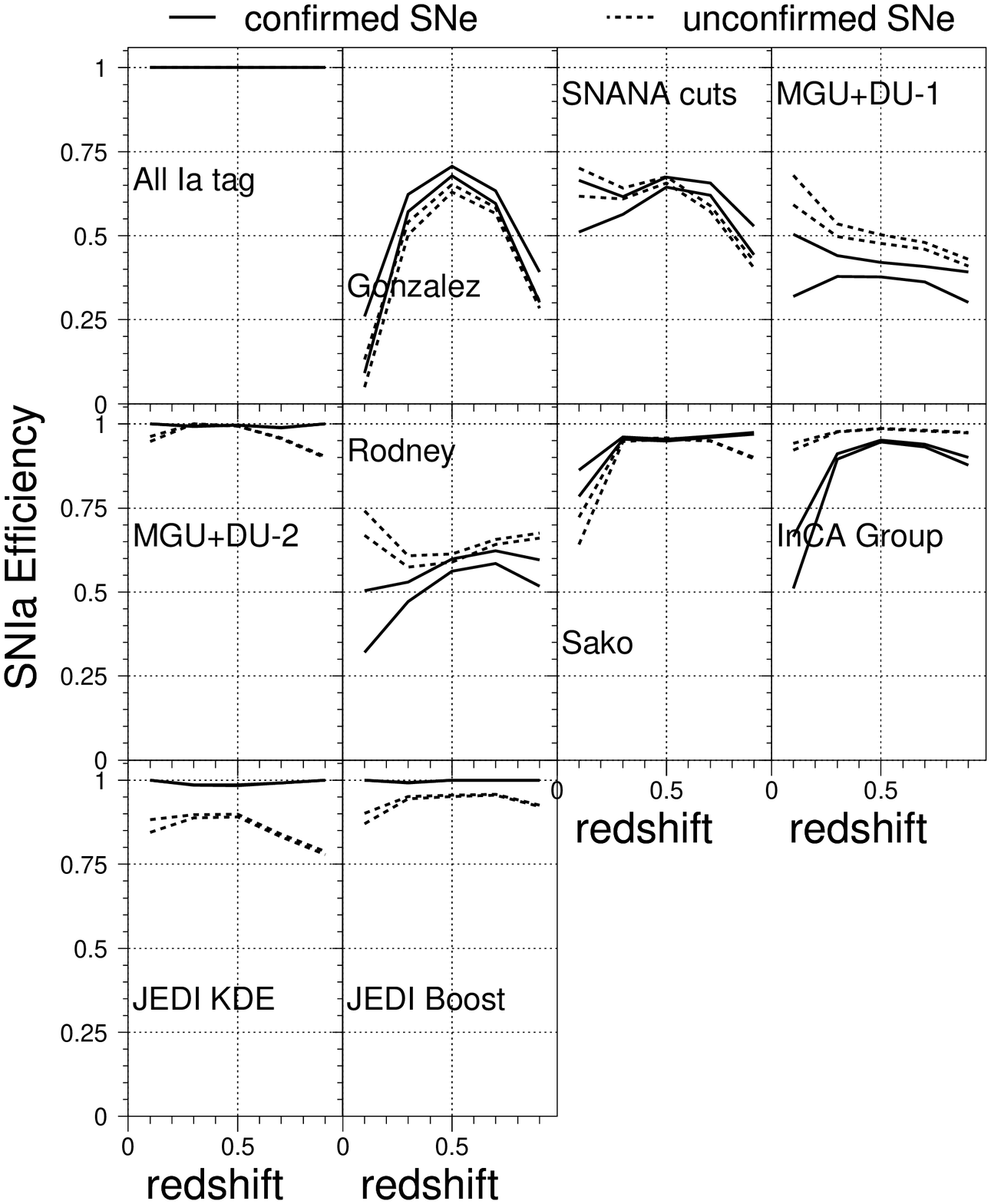}  
\plottwo{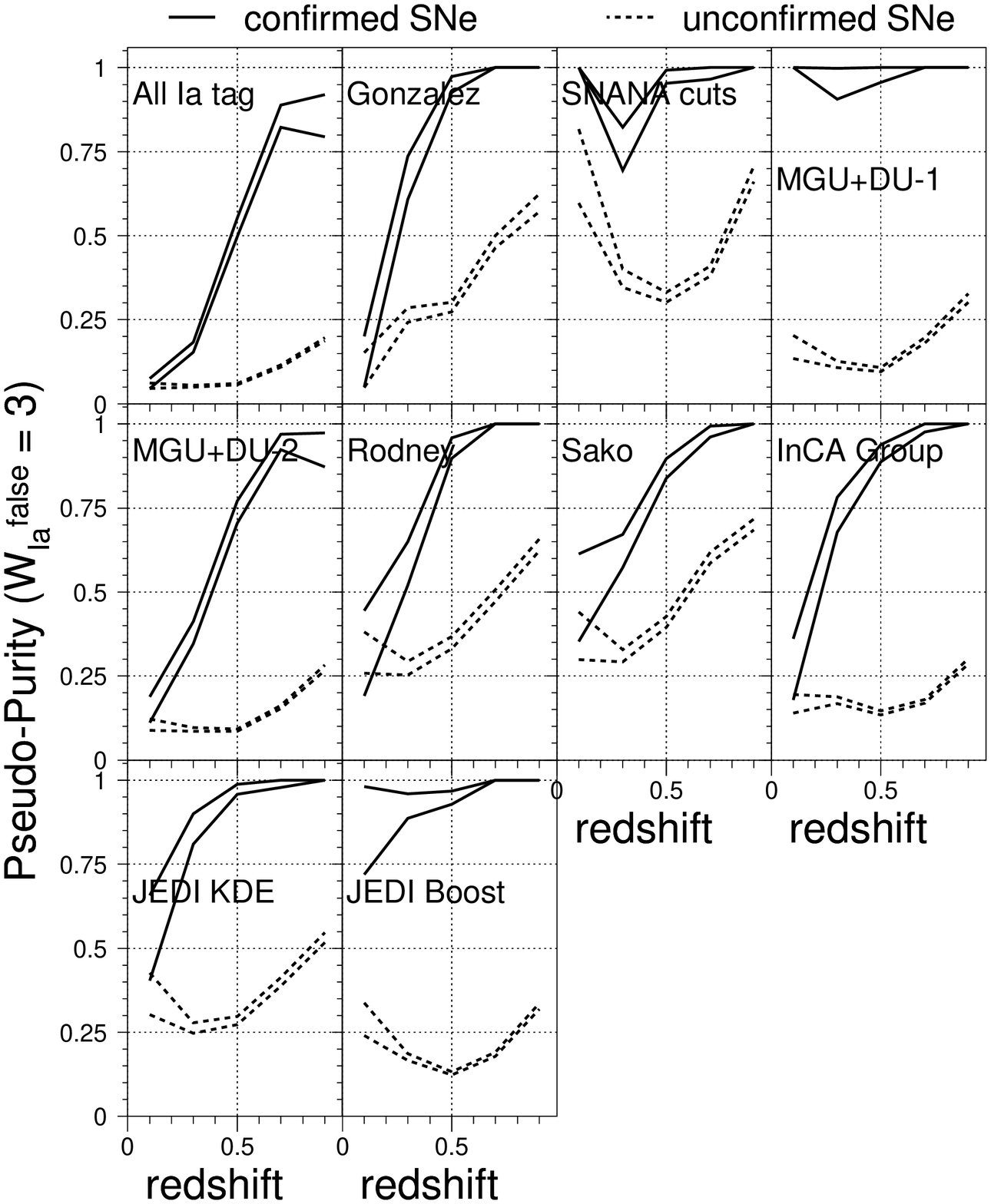}{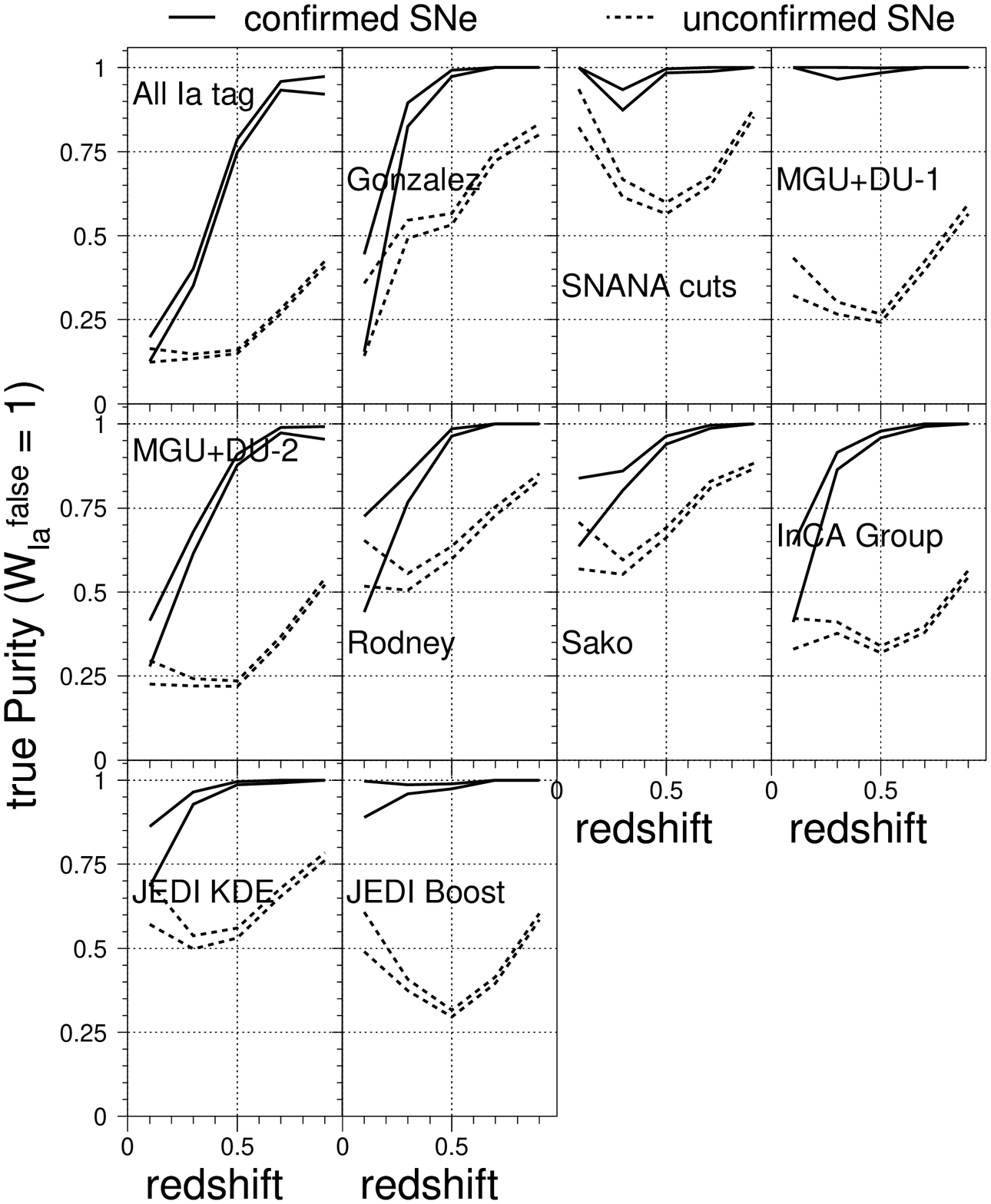}  
  \caption{
	For each participant in the \SNPCCnoHOSTZ,	
	results vs. redshift are shown for $\FoMIa$, $\effspec$,
	pseudopurity ($\PurityIa$), and the true purity
	($\PurityIa$ with $\wIafalse=1$).
	{\curvedefSNIa}
      }
  \label{fig:nohostz_resultsIa}
\end{figure*}

\begin{figure*}
\centering
\epsscale{0.9}
\plottwo{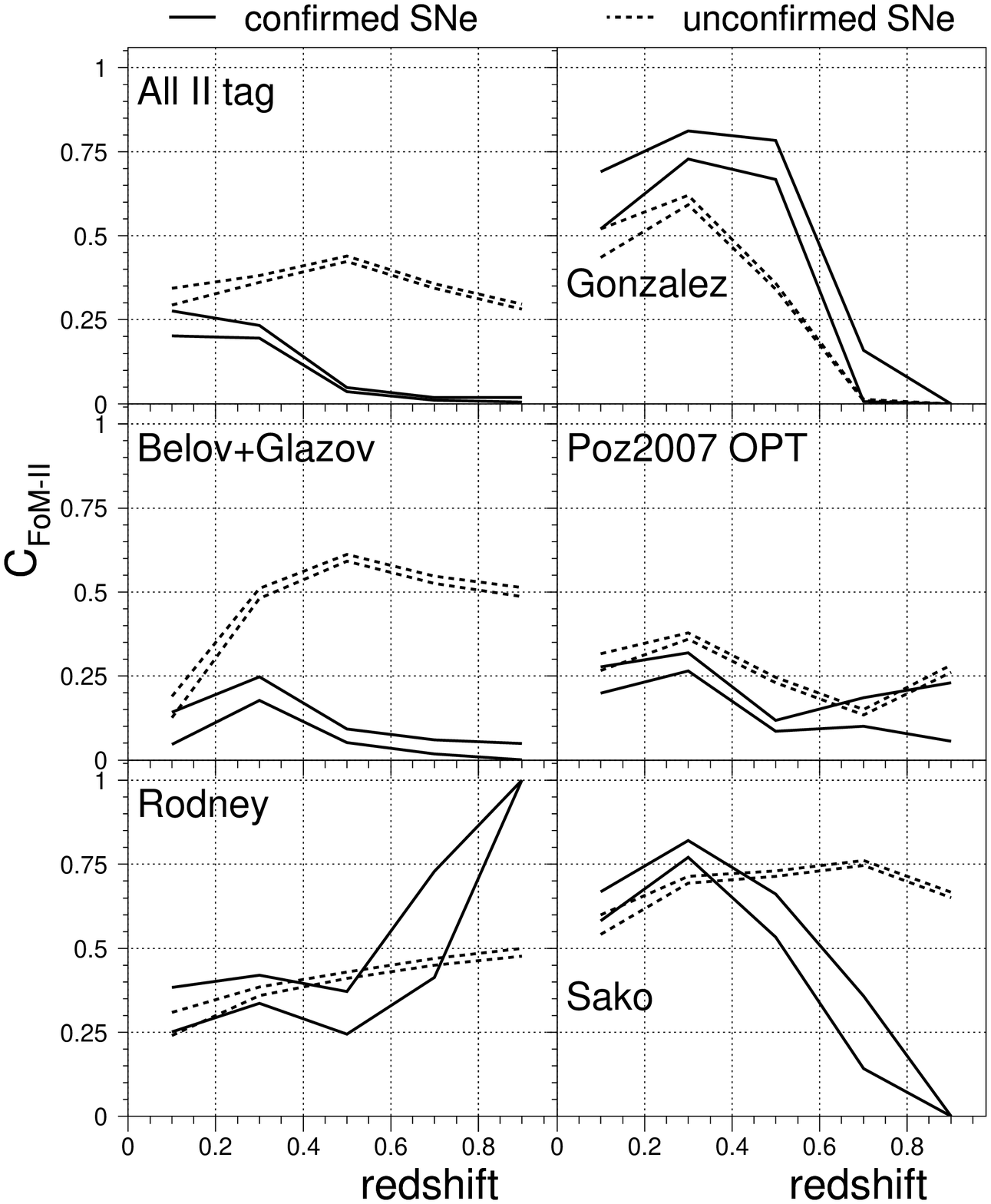}{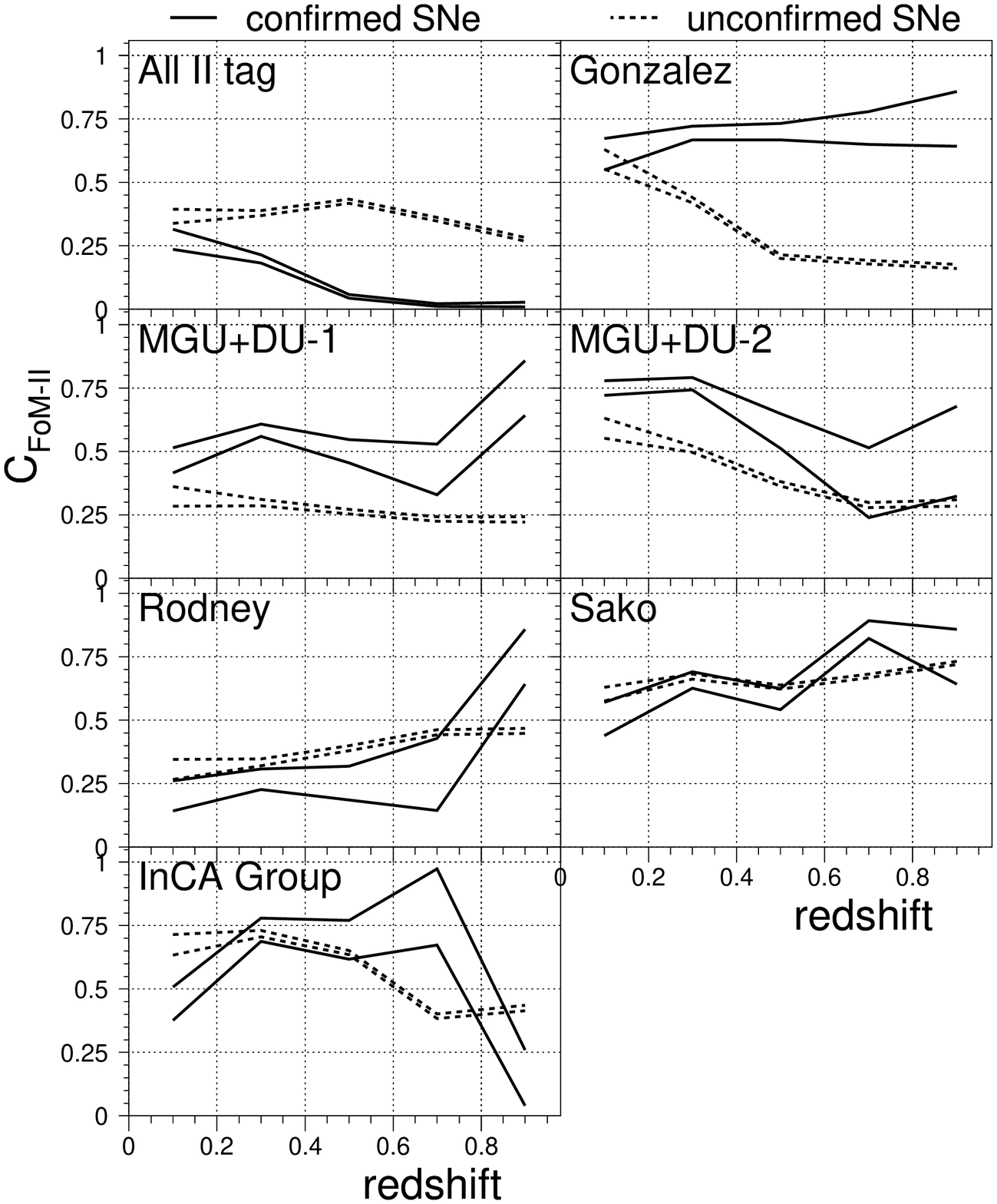}  
  \caption{
    Figure of merit versus redshift for type II classifications
    from the \SNPCCHOSTZ\ (left) and from the \SNPCCnoHOSTZ\ (right).
    {\curvedefSNII}
      }
  \label{fig:resultsII}
\end{figure*}

The results from the SN~Ia evaluations 
(\S\ref{sec:eval}) are shown in 
Figures~\ref{fig:hostz_resultsIa}-\ref{fig:nohostz_resultsIa},
corresponding to the challenges with and without 
host-galaxy \photoz\ information.
As a function of the true (generated) redshift, we have plotted
the figure-of-merit quantity $\FoMIa$ (Eq.~\ref{eq:FoM}),
efficiency ($\effIa$), pseudopurity ($\PurityIa$),
and true purity.
For each variable, the redshift dependence is shown separately 
for the \specy\ confirmed subset (solid) and the
unconfirmed SNe (dashed).
The label on each panel indicates the name of the participant
or group. 
The first panel labeled ``All Ia tag''
is an arbitrary reference in which every SN has been
tagged as an SN~Ia, thereby ensuring 100\% efficiency.
The corresponding results for type II classifications
are shown in Fig.~\ref{fig:resultsII}.

For the SN~Ia classifications,
the most notable trend in all of the entries
is that the figure of merit ($\FoMIa$)
is significantly worse for the unconfirmed sample
than for the \specy\  confirmed subset.
Depending on the redshift, the confirmed-unconfirmed
differences vary by tens of percent to nearly an order
of magnitude.
Several methods show improving $\FoMIa$ with redshift. 
We see this trend for the \specy\ confirmed ``All Ia'' 
entry because at high redshift anything bright enough to
obtain a spectrum is likely to be an SN~Ia.

For the unconfirmed SN subset, the largest $\FoMIa$ value in any 
redshift bin is about 0.6, but these entries show at least a
factor-of-2 variation in $\FoMIa$ as a function of redshift.
The most stable figure of merit versus redshift (for unconfirmed SNe)
has $\FoMIa = 0.3$ -- 0.45 at all redshifts.
The largest variation is $0.1 < \FoMIa < 0.6$.

In spite of the caveats about trying to determine
the best method in this first \SNPCC,
here we carefully examine the $\FoMIa$ for the unconfirmed sample 
in the \SNPCCHOSTZ\ (Fig.~\ref{fig:hostz_resultsIa}).
The entry with the highest average
figure of merit (Sako) has an average
SN~Ia efficiency of \BESTEFF\ and an average
SN~Ia purity (i.e., $\wIafalse=1$) of \BESTPURITY.
However, comparing the best figure of merit (vs. redshift)
for each strategy shows that three strategies yield similar results:
selection cuts, Bayesian probabilities and statistical
inference. The remaining Hubble-diagram strategy is 
somewhat worse at low and high redshifts.
Among the entries for a given strategy
there is a large variation in the figure of merit,
suggesting that the optimum has not been achieved.
For participants who applied the same method to both
the \SNPCCHOSTZ\ and the \SNPCCnoHOSTZ, 
the average $\FoMIa$ was smaller for the \SNPCCnoHOSTZ\ 
by as little as 6\% (Sako and JEDI-KDE) 
and by as much as a factor of 2.

\begin{figure*}
\centering
\epsscale{.9}
\plottwo{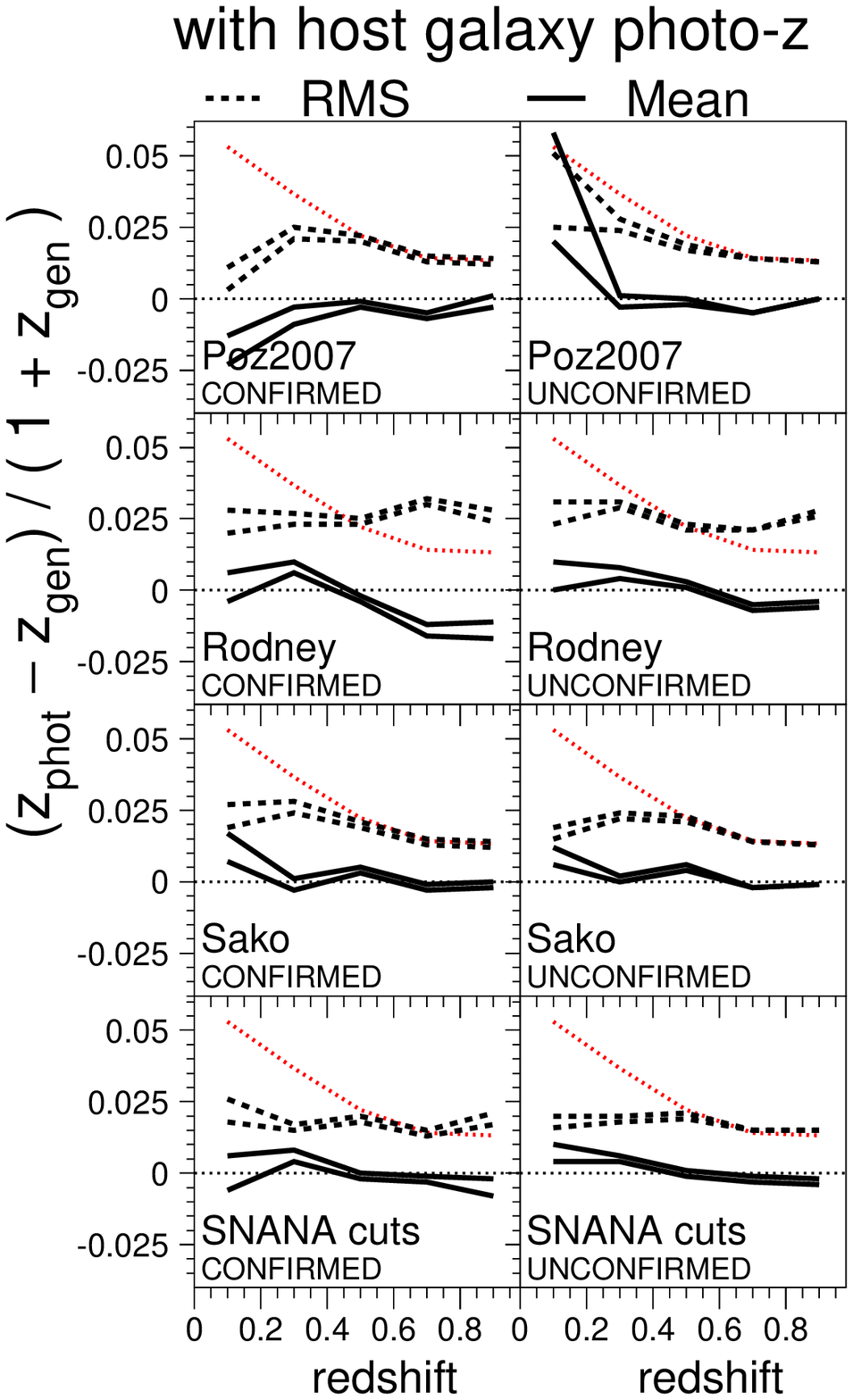}{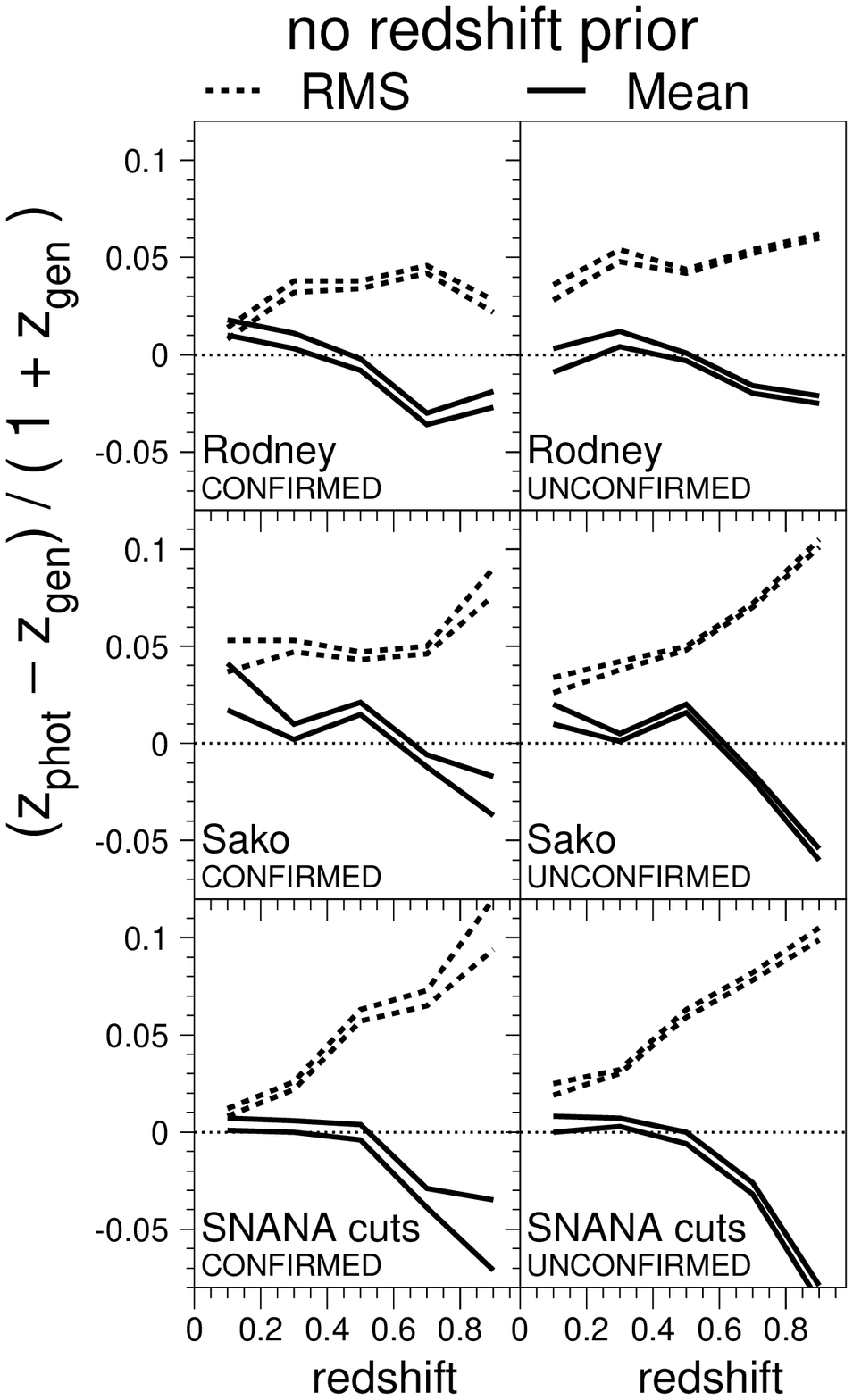} 
  \caption{
	Photo-$z$ residuals $\DZPHOT$ vs. redshift for the
	\SNPCCHOSTZ\ (left) and \SNPCCnoHOSTZ\ (right).
	The mean residual is shown the by solid curves,
	and the rms by the dashed curves.
	The dotted curves (same in each \SNPCCHOSTZ\ panel)
	show the rms for the host-galaxy \photoz.
	Note that the vertical scales are different
	for the left and right plots.
	}
  \label{fig:photoz_resid}
\end{figure*}

The \photoz\ residuals are shown in  Fig.~\ref{fig:photoz_resid}
for those entries that include \photoz\ estimates.
Here we show residuals only for true SNe~Ia that have been
correctly typed as an SN~Ia.
When the host-galaxy \photoz\ is available, 
the supernova light curve improves the \photoz\ precision 
for redshifts up to about 0.4.
For the \SNPCCnoHOSTZ, the bias and scatter of the residuals
is significantly larger than for the \SNPCCHOSTZ.

\bigskip

After evaluating the classification results and algorithms,
two notable problems were identified in the implementations.
First, the \specy\ confirmed subset was generally treated as 
a random subset, which it clearly is not 
(\S\ref{subsec:sim_spec_subset}). 
The magnitude-limited selection of \spec\ targets
resulted in the selection of brighter objects in the
training subset. In principle, the brighter objects
in the training subset should be re-simulated at higher
redshifts so that classification algorithms can
be trained on more distant (dimmer) objects for which
spectra cannot be obtained.

The second general problem is that several
entries did not use all available information from 
the  light curves (most notably, ignoring colors), 
or effectively added noise to the information.
The latter was mainly an artifact from a very poor
determination of the epoch of maximum brightness.
Specific details of these problems 
are given in Appendix~\ref{app:methods}.


 \clearpage  

 \section{Updated Simulations}
 \label{sec:sim_update}

While we have no plans for another competition-style challenge, 
we have released updated simulated samples as a public resource 
for the development of photometric SN classification and \photoz\ 
estimators.\footnote{\wwwSIMGEN}
For these updated samples we have fixed the known bugs
(\S\ref{subsec:sim_bugs}), made some improvements,
provided additional samples corresponding to the
LSST \citep{LSSTSB09} and \SDSS\ surveys, 
and included the answer keys
giving the generated type and other parameters for each SN.
The answer keys will allow developers to study
different \specy\ confirmed training subsets,
and to evaluate their own analysis.

The updated simulations have two main improvements related
to the generation of SNe~Ia. 
The first improvement is a more realistic modeling of 
color variations based on recent results from
\citet{Guy2010}. The newly measured  variation is about 
0.05~mag (Gaussian sigma) in the ultraviolet wavelength region 
and $\sim 0.02$~mag  in the other wavelength regions.
These variations are significantly smaller than 
what was used in the \SNPCC, where 
an independent variation per passband was drawn randomly
from a Gaussian distribution with $\sigma_m =0.09$~mag.
To obtain a reasonable Hubble scatter in the updated 
simulations, a 0.12~mag random Gaussian smearing is 
added coherently to all epochs and passbands.
The second improvement is to use more realistic
distributions of color and stretch ($x_1$ parameter)
for the SNe~Ia generated with the \SALTII\ model.
These distributions include more realistic tails
corresponding to dimmer SNe, 
resulting in fewer \SALTII-generated SNe~Ia satisfying
the loose selection criteria.
The sample sizes generated from the \mlcs\ and \SALTII\ models
are thus very similar, in contrast to the larger \SALTII\ sample
in the \SNPCC\ (\S\ref{subsec:sim_bugs}).

 \section{Conclusion}
 \label{sec:end}

We have presented results from the SN classification challenge
that finished 2010 June 1. 
Among the four basic strategies that were
used in the \SNPCC\ (\S\ref{sec:take_challenge}),
three strategies show comparable results for the
entries with the highest figure of merit.
Therefore no particular strategy was notably superior.
For all of the entries, the classification performance was 
significantly better for the \spec\ training subset
than for the unconfirmed sample.
The degraded performance on the unconfirmed sample was
in part due to participants not accounting for the
bias in the \spec\ training sample.

There is a large variation in the figure of merit 
and  therefore we urge caution in using these evaluations 
to determine the best method. 
The quality of each implementation varies significantly 
between participants (Appendix~\ref{app:methods}) 
and therefore some improvements
are needed before drawing more clear conclusions. 
While this article signifies the end of the \SNPCC, 
we consider this effort to be the start of a new era
for developing classification methods with significantly 
improved simulation tools.
The results from  this \SNPCC\ may serve as a reference
to assess future progress from using improved algorithms
and improved simulations. 
As described in \S\ref{sec:sim_update}, 
these updated simulations, along with the answer keys
giving the true type for each SN, are publicly available.

While the optimal classification algorithm can in principle
be optimized after a survey has completed,
it is advantageous to define the necessary \spec\ training 
sample before a survey has started. 
In particular, is a magnitude-limited training sample 
adequate (i.e., as used in this {\SNPCC}), or is a less
biased training sample needed?
The latter sample is clearly more desirable for
training classification algorithms,
but this strategy results in fewer \specy\ confirmed SNe~Ia.
As described in \S\ref{sec:sim_update}, 
this issue can be investigated more thoroughly
by defining arbitrary \spec\ training subsets
for the publicly available simulated samples.

To optimize the use of a magnitude-limited sample,
we suggest another strategy that was not tried by any 
of the participants. In principle the \specy\ confirmed 
\nonIa\ sample can be used to simulate \nonIa\ SNe at higher 
redshifts to obtain an extended training sample for
the classification algorithms. In contrast to an
ideal unbiased \spec\ sample however, 
this simulation strategy does not account 
for changes in the relative rates with redshift.

The figure of merit used in this challenge (\S\ref{sec:eval})
allows for a quantitative comparison between methods,
but does not quantify the impact of photometric classification 
on the inference of cosmological parameters.
Therefore, an important next step in using these simulations 
is to carry out a full analysis that includes the determination 
of cosmological parameters from a Hubble diagram.

\bigskip

We are grateful to the 
Carnegie Supernova Project (CSP), 
Sloan Digital Sky Survey-II ({\SDSS}) and 
Supernova Legacy Survey  collaborations
for providing unpublished \specy\ confirmed \nonIa\ 
light curves that are critical to this work.
Funding for the creation and distribution of the SDSS and SDSS-II
has been provided by the Alfred P. Sloan Foundation,
the participating institutions,
the National Science Foundation,
the U.S. Department of Energy,
the National Aeronautics and Space Administration,
the Japanese Monbukagakusho,
the Max Planck Society, and 
the Higher Education Funding Council for England.
The CSP has been supported by the National Science
Foundation  under grant AST--0306969.

\bigskip

\clearpage
\begin{appendix}

\section{Classification Methods from {\SNPCC} Participants}
\label{app:methods}

{\bf Belov \& Glazov.}---
For each SN from the challenge, the public \SNANA\ simulation 
was used to generate simulated SNe at the same epochs as the 
challenge SN.
The epoch of peak brightness ($t_0$) was estimated to be
18 days after the first $g$-band measurement,
thereby taking advantage of a bug in the \SNPCC\
(Table~\ref{tb:bugs}). However, this estimate of $t_0$
did not account for the redshift or stretch.
Types Ia, \Ibc, \IIP, \IIn,  and \IIL\ were generated,
and the \nonIa\ were based solely on the publicly available 
Nugent SED templates.
The classification was then based on the minimum $\chi^2$ 
between the challenge SN and the {\SNANA}-generated SNe. 
\SNPCC\ SNe with large minimum $\chi^2$ were rejected.

{\bf Gonzalez.}---
SN~Ia identification used the \SIFTO\ light-curve fitter
\citep{Conley2008} that was developed by the SNLS.
This fitting program was modified to include the 
redshift as a free parameter \citep{Sulli2006}.
The fitted values of the color, stretch and $\chi^2$
were used to determine if a candidate SN was a type Ia,
but these values were not used to classify a \nonIa\ subtype.
Type \IIP\ identification was based on a postmaximum linear fit 
(in magnitudes per day) in each band. 
From the training sample, the resulting slope
in each passband was used to define a probability space.

{\bf InCA.}---
This method labeled supernovae by performing classification 
on a lower-dimensional representation of the SN light 
curves without relying on the use of templates or measured 
physical parameters such as stretch and color. 
Specifically, the diffusion map approach to 
nonlinear dimensionality reduction  \citep{JR09} was utilized.  
Using these lower-dimensional objects, well-established methods
for classification were implemented to estimate the type for 
each unknown SN.

The diffusion map was based on a pairwise distance measure over 
all of the observed light curves and bands. This distance matrix 
was then smoothed and transformed into diffusion space, 
providing the dimensionality reduction and possibly illuminating 
structure hidden in the original representation.

To compute these distances, a regression spline was first fit to each 
SN light curve in each filter. This allowed each SN to be represented
as fluxes (and errors) on 1-day intervals. The time axis was shifted 
so that the observer-frame time of peak $r$-band brightness was the 
same for each SN and the fluxes were normalized so that each SN has
the same maximum $r$ band flux.  These steps were performed to 
ensure that the subsequent steps capture differences in the 
shapes and colors of each light curve.
A potential weakness, however, was that using the 
observer-frame $r$-band as a reference does not match 
the peak colors and epochs in the rest frame.
Using the normalized spline fit from each band of each light curve, 
the distance between SNe $i$ and $j$ in band $b$ was defined as
\begin{equation}
\label{eqn:distband}
d_{ij}^b = \frac{1}{\Delta T_b} 
   \left(\sum_e [F^b_{i,e}-F^b_{j,e}]^2/ 
    [(\sigma^b_{i,e})^2 + (\sigma^b_{j,e})^2]\right)^{1/2}~~,
\end{equation}
where $\Delta T_b$ is the amount of overlap time (days) 
between the two SN light curves,
$F^b_{i,e}$ is the spline-fitted flux of SN $i$ in band $b$ at epoch $e$, 
$\sigma^b_{i,e}$ is the fitted error, 
and the epoch index $e$ runs over the overlapping time bins.  
The distance between each pair of SNe was constructed as the 
linear (not quadratic) sum of $d_{ij}^b$ over bands,  
$d_{ij} = \sum_b d_{ij}^b.$
Next, the distance matrix $d_{ij}$ was smoothed and 
transformed into an $m$-dimensional representation of each SN that 
best preserves the relationships between each pair of SNe in the 
context of a diffusion process over the data.  
This lower-dimensional representation was used (with $m$=50) 
in conjunction with the random forest  classification  method 
(Breiman 2001) to estimate the 
type of each SN based on the set of training SNe.


{\bf JEDI KDE.}---
The light curve for each filter was fit to a modified 
$\Gamma$-distribution function with five parameters. 
The four filters and redshift resulted in 21 parameters. 
A Gaussian was constructed around each 21-parameter point 
with a variance related to the density of points in its vicinity. 
The sum of these Gaussians over the spectroscopic training 
subset constitutes the Kernel Density Estimator (KDE). 
A relative probability of being a type Ia or \nonIa\ SN 
for any set of 21-parameters was obtained from the 
Ia and non-Ia KDEs. 
A selection cut on the KDE probabilities 
was used to make classifications.

{\bf JEDI boost.}---
This method used a supervised learning algorithm for classifying 
high-dimensional, nonlinear data \citep{Hastie2009}.
The idea was to combine decisions from a group of
weak classifiers to make a more informed decision.
This algorithm used the 21 parameters from the light-curve fit, 
plus the two KDE probabilities. The tree depth was 3,
and the number of trees was 2000.

{\bf JEDI-Hubble.}---
The \spec\ training subset was used to construct a Hubble diagram
and a two-dimensional KDE was constructed for the 
type Ia and \nonIa\ SNe. This method is similar to that
of the Portsmouth-Hubble entry which used $\chi^2$ statistics
instead of a KDE.

{\bf JEDI combo.}---
This method combined the KDE probabilities from the 
JEDI-Boost and JEDI-Hubble methods.


{\bf MGU+DU-1.}--- 
The \spec\ training  subset was used to estimate light-curve 
slopes (mag/day) in each filter in four separate observer-frame 
regions relative to the epoch of peak brightness: 
$-25$ to $-1$, 1 to 25, 20 to 75 and 60 to 110 days. 
Redshift information was not used to translate these slopes 
into the rest frame, and each filter was treated independently 
so that color information was not used either.
The slopes for each SN were then compared with the expected slopes 
for each SN type using a 
``difference boosting neural network'' (DBNN; \citet{Philip2000}).
If the same class was predicted in three or more filters,
that class was used. In case of a tie, 
where two classes were each predicted by two filters, 
the product of the confidences was used to determine the class, 
with the one with the higher product winning.
If there were no predictions,  or if several classes were 
predicted by one filter each, the SN was rejected.

{\bf MGU+DU-2.}---
This method was nearly the same as that for MGU+DU-1, 
except that a machine learning method called 
random forests \citep{Breiman2001}
was used to determine the predictive model.


{\bf Portsmouth $\chi^2$.}---
This classification was based on the $r$-band $\chi^2$ from 
\SALTII\ light-curve fits \citep{Guy07}. 
The $\chi^2_r$ cut was determined by optimizing the $\FoMIa$
on the training sample using the 
false discovery rate statistic \citep{Miller2001}.
The only selection requirement was that the
\SALTII\ fit does not fail or return pathological values.

{\bf Portsmouth-Hubble.}---
For the \specy\ confirmed subset, a Hubble diagram (HD)
was generated by the \SALTII\ light-curve fits.
This HD was then fit to a fourth-order polynomial, 
resulting in an  expected HD curve that has no assumptions 
about cosmological parameters. 
For the unconfirmed sample, a $\chi^2$ was computed 
for each SN based on the proximity of the distance modulus 
to the expected HD curve.  The $r$-band $\chi^2$ from the
previous entry was not used.


{\bf Poz2007 RAW.}---
The SN automated Bayesian classifier (SN-ABC; \citet{Poz2007a_classify})
was used without any modifications.
The light-curve templates included one SN~Ia 
(no stretch or color dependence) and the 
\IIP\ SED template from Nugent.

{\bf Poz2007 OPT.}---
SN-ABC was used as in the previous entry, and included 
selection cuts based on optimizing the figure of merit
(\S\ref{sec:eval}) for the \specy\ confirmed subset.

{\bf Rodney.}---
The method of ``Supernova Ontology with Fuzzy Templates''
(SOFT; \citet{Rodney2009,Rodney2010})  was used with three
significant adjustments.
First, the \specy\ confirmed subset was used to define a 
redshift-dependent probability for each class.
Next, instead of fixing the extinction parameter $R_V$,
it was allowed to take three discrete values: 1.3, 2.2, or 3.1.
Finally, the host-galaxy \photoz\ was included as a prior 
for the \SNPCCHOSTZ.	
To reduce the processing time without dramatically affecting 
the results, the \spec\ training set from the challenge
was used to reduce the SOFT template library 
from $\sim 40$ templates down to 20.

\newcommand{\MSonly}{$^\dagger$}
\newcommand{\SNPCCovp}{$^\star$}

{\bf Sako.}---
This entry used an improved version of the method used 
to classify objects during the \SDSS\ SN Survey \citep{Sako08}.
A $\chi^2$ was computed between the observed photometry
and each SN from a large set of templates that included
SN~Ia and \nonIa\ light-curve models.
For the SN~Ia models there were 5 parameters defining
a grid of 45 million templates:
1) redshift, 
2) rest-frame $V$-band extinction, 
3) time of maximum light in $B$ band,
4) shape-luminosity parameter $\Delta m_{15}$ \citep{Phillips_93}, and
5) distance modulus.  
Flat priors were assumed for all parameters except when 
the host-galaxy redshift was available.
The \nonIa\ templates were based on \specy\ confirmed
\SDSS\ SNe including 
type \Ibc\ 
(2005hl{\MSonly}, 2005hm{\SNPCCovp}, 2006fo{\SNPCCovp}, and 2006jo{\SNPCCovp}) 
and type II 
(2004hx{\SNPCCovp}, 2005lc{\MSonly}, 
 2005gi{\SNPCCovp}, and  2006jl{\SNPCCovp}).
The star (dagger) superscript indicates that this SN was (was not) 
used in the \SNPCC\ (see Table~\ref{tb:nonIa_list}).
Although the choice and development of these templates were 
completely independent of the \SNPCC, this method clearly
had an advantage in using a few of the same templates that
were used in the \SNPCC.

%
%

SNe with large $\chi^2$ were rejected. 
The final SN classification was based on the largest 
Bayesian probability among the calculated probabilities 
to be a type Ia, \Ibc\ or  II.
This algorithm is similar to the one presented in \citet{Poz2007a_classify} 
except that \nonIa\ SNe were classified into subtypes \Ibc\ and II 
using an extended set of templates, 
the distance modulus was allowed to vary 
(instead of being computed from the SN {\photoz} 
and an assumed cosmology)
and the shape parameter was allowed to vary
for SN~Ia light curve models.

{\bf \SNANA\ Cuts.}---
Two of the challenge organizers (SK \& RK) created a submission
using the {\SNANA}-\mlcs\ light-curve fitter along with
selection cuts that were guessed long before the \SNPCC.
We did not optimize the cuts, or use our in-depth knowledge 
of how the \SNPCC\ was generated. The primary cut required
that the \mlcs\ light-curve fit probability be above 10\%.
The other selection requirements were 
1) at least one measurement before the epoch of peak brightness
and another 10 days later in the rest frame,
2) maximum S/N$>10$ and 
3) two additional filters with maximum S/N$>5$.
The \photoz\ estimates used the method described	
in \citet{K10photoz}.


\end{appendix}


\clearpage
\bibliographystyle{apj}
\bibliography{SNchallenge_results}

  \end{document}